%% file: main.tex
\documentclass[conference]{IEEEtran}

\IEEEoverridecommandlockouts

\usepackage{cite}

\usepackage[utf8]{inputenc}
\usepackage{amsmath}
\usepackage{amssymb}
\usepackage{ebproof}
\usepackage{stmaryrd}
\usepackage[breakable]{tcolorbox}
\usepackage{xspace}
\usepackage{listings}

\usepackage{textcomp}
\definecolor{lgreen}{RGB}{248,255,248}
\definecolor{dgreen}{RGB}{0,128,0}
\definecolor{titlecol}{RGB}{170,199,250}
\definecolor{dred}{RGB}{200,0,0}
\definecolor{lyellow}{RGB}{255,255,40}

\newtheorem{definition}{Definition}
\newtheorem{example}{Example}

\title{Modular System Synthesis}
\author{\IEEEauthorblockN{Kanghee Park \qquad 
Keith J.C. Johnson \qquad 
Loris D'Antoni \qquad 
Thomas Reps}
\IEEEauthorblockA{\textit{University of Wisconsin--Madison} \\
Madison, USA \\
\{khpark, keithj, loris, reps\}@cs.wisc.edu
}
}

\newcommand{\mypar}[1]{\vspace{1mm}\noindent\textit{#1.}}

\newcommand{\rone}{(\emph{i})\xspace}
\newcommand{\rtwo}{(\emph{ii})\xspace}
\newcommand{\rthree}{(\emph{iii})\xspace}

\newcommand{\framework}{\textsc{MoSS}\xspace}
\newcommand{\tool}{\textsc{MoSSKit}\xspace}

\newcommand{\spyro}{\textsc{spyro}\xspace}
\newcommand{\sketch}{\textsc{sketch}\xspace}
\newcommand{\jlibsketch}{\textsc{JLibSketch}\xspace}

\newcommand{\exname}[1]{\mathtt{#1}}

\newcommand{\modname}[1]{\texttt{#1}}
\newcommand{\modlist}{\modname{List}\xspace}
\newcommand{\modstack}{\modname{Stack}\xspace}
\newcommand{\modqueue}{\modname{Queue}\xspace}
\newcommand{\modticket}{\modname{TicketVendor}\xspace}

\newcommand{\modarraylist}{\modname{ArrayList}\xspace}
\newcommand{\modhm}{\modname{HashMap}\xspace}
\newcommand{\modhs}{\modname{HashSet}\xspace}
\newcommand{\modts}{\modname{TreeSet}\xspace}
\newcommand{\modhmone}{\modname{HashMap1}\xspace}
\newcommand{\modhmtwo}{\modname{HashMap2}\xspace}
\newcommand{\modkafka}{\modname{Kafka}\xspace}

\newcommand{\error}{err}

\newcommand{\arrayget}{\exname{get}}
\newcommand{\arrayset}{\exname{set}}
\newcommand{\arrayec}{\exname{ensureCapacity}}

\newcommand{\mapempty}{\exname{emptyMap}}
\newcommand{\mapput}{\exname{put}}
\newcommand{\mapget}{\exname{get}}

\newcommand{\vark}{k}
\newcommand{\vari}{i}
\newcommand{\varj}{j}
\newcommand{\varl}{l}
\newcommand{\varn}{n}
\newcommand{\varm}{m}

\newcommand{\varval}{v}

\newcommand{\kone}{k_1}
\newcommand{\ktwo}{k_2}

\newcommand{\inttype}{\modname{int}}

\newcommand{\ite}{\exname{ite}}

\newcommand{\lnil}{\exname{nil}}
\newcommand{\lcons}{\exname{cons}}
\newcommand{\lhead}{\exname{head}}
\newcommand{\ltail}{\exname{tail}}
\newcommand{\lsnoc}{\exname{snoc}}

\newcommand{\lsize}{\exname{sizeL}} % shortened: size_l
\newcommand{\isemptylist}{\exname{isEmptyL}} % shortened: is_empty_l

\newcommand{\emptystack}{\exname{emptyS}} % shortened: empty_s
\newcommand{\push}{\exname{push}}
\newcommand{\pop}{\exname{pop}}
\newcommand{\sttop}{\exname{top}}
\newcommand{\stsize}{\exname{sizeS}} % shortened: size_s
\newcommand{\isemptystack}{\exname{isEmptyS}} % shortened: is_empty_s

\newcommand{\emptyqueue}{\exname{emptyQ}} % shortened: empty_q
\newcommand{\enq}{\exname{enq}} % shortened: enqueue
\newcommand{\deq}{\exname{deq}} % shortened: dequeue
\newcommand{\front}{\exname{front}}
\newcommand{\isemptyqueue}{\exname{isEmptyQ}} % shortened: is_empty_q
\newcommand{\qsize}{\exname{sizeQ}} % shortened: size_q

\newcommand{\preparesales}{\exname{prepSales}} % shortened: prepare_sales
\newcommand{\issueticket}{\exname{issueTicket}} % shortened: issue_ticket
\newcommand{\reserveticket}{\exname{resTicket}} % shortened: reserve_ticket
\newcommand{\soldout}{\exname{soldOut}} % shortened: sold_out
\newcommand{\numticket}{\exname{numTicketsRem}} % shortened: num_tickets_remaining
\newcommand{\numwaiting}{\exname{numWaiting}} % shortened: num_waiting

\newcommand{\tru}{\top}
\newcommand{\fls}{\bot}

\newcommand{\searchspace}{S}
\newcommand{\propertyspace}{\Phi}
\newcommand{\signatures}{\Sigma}

\newcommand{\PL}{\mathcal{P}}
\newcommand{\MI}{\textit{MI}}
\newcommand{\PF}{\textit{PF}}
\newcommand{\iss}[1]{\varphi^{#1}_{\textit{imp}}}
\newcommand{\ias}[1]{\varphi^{#1}_{\textit{sem}}}
\newcommand{\iasprop}{\alpha}
\newcommand{\lang}{\mathcal{L}}

\newcommand{\phil}{\varphi}
\newcommand{\lproperty}{$\lang$-property\xspace}

\newcommand{\lconjunctions}{$\lang$-conjunctions\xspace}
\newcommand{\lproperties}{$\lang$-properties\xspace}

\usepackage{csquotes}
\usepackage{framed}
\usepackage{arydshln}
\usepackage{afterpage}
\usepackage{enumitem}
\usepackage[caption=false]{subfig}
\usepackage{tikz}

\newcommand{\blackcircle}[1]{\tikz[baseline=(char.base)]{
    \node[shape=circle,draw,inner sep=2pt,fill=black,text=white] (char) {\scriptsize #1};}}
\newcommand{\callout}[1]{\blackcircle{#1}}

\setlist[itemize]{align=parleft,left=0pt..1em, topsep=2pt, itemsep= 0pt}
\setlist[description]{topsep=2pt}

\begin{document}

\maketitle

\input{abstract}
\input{1-introduction} 
\input{2-illustrative-example}
\input{3-modular-system-design}
\input{4-synthesis}
\input{5-case-study}
\input{6-related-work}
\input{7-conclusion}

\bibliographystyle{abbrv}
\bibliography{main}

\clearpage
\appendix
\input{8-appendix-detailed-case-study}

\input{9-appendix-code}

\end{document}

%% file: abstract.tex
\begin{abstract}
% Synthesis currently only scales to small programs 

% If we want synthesis to scale to larger programs, we need a way to enable compositionality---i.e., solving larger in terms of smaller synthesis problems

% Enabling compositionality requires layering problems so that solving problems at a layer only depends on that layer and on the solutions obtained at lower layers

% To do so, we need to solve two problems: 1) have a way to feed the results at lower layers to the layer we are trying to solve, 2) solve the individual synthesis problems.

% In other words, our goal is to support synthesis via modular software design with information hiding.

% This paper presents an approach that enables this kind of compositionality

% Contribution 1: The SemGuS framework as a way to express and solve problems at individual layers

% Contribution 2: A way to translate solution to lower-layer problems to parts of SemGuS problems at higher layers (key idea synthesizing specifications)

% 1. State the problem
% Program synthesis aims to ease the burden on programmers by automatically creating a program that satisfies a specification provided by a user.
% % 
% Unfortunately, existing tools do not scale to large programming problems.

%\hspace*{1.5ex}
% 2. Say why it’s an interesting problem
% 3. Say what your solution achieves
This paper describes a way to improve the scalability of program synthesis by exploiting \emph{modularity}:
larger programs are synthesized from smaller programs.
The key issue is to make each ``larger-created-from-smaller'' synthesis sub-problem be of a similar nature, so that the kind of synthesis sub-problem that needs to be solved---and the size of each search space---has roughly the same character at each level.
% 4. Say what follows from your solution
This work holds promise for creating program-synthesis tools that have far greater capabilities than currently available tools, and opens new avenues for synthesis research:
how synthesis tools should support modular system design, and how synthesis applications can best exploit such capabilities.

% \loris{as rewritten by chatgpt}
% The aim of program synthesis is to simplify the programming process by finding solutions for special cases automatically, making it possible for non-programmers to create programs simply by specifying the desired outcome. However, existing tools face a scalability issue when applied to large programming problems.

% This paper presents a solution to this scalability issue by introducing a modular approach to program synthesis, where large programs are constructed from smaller programs. The challenge is to ensure that the synthesis problems at each level are of a similar nature, leading to a consistent structure for the synthesis problem and search space.

% By employing this modular approach, program synthesis tools will have enhanced capabilities compared to current tools, and open up new avenues for synthesis research, including the best ways to support modular system design and utilize the capabilities of synthesis tools.
\end{abstract}

%% file: 1-introduction.tex
\section{Introduction}
\label{Se:Introduction}

% C: Context
%
In program synthesis, the goal is to automatically (or semi-automatically) create programs that match high-level intents provided by a user---e.g., logical specifications or input-output examples.
%
% % G:Gap
To date, however, synthesis tools cannot contend with large programs because they require synthesizing (or at least reasoning about) a program in its entirety.

The obvious direction is to try to exploit \emph{compositionality} and synthesize larger programs by having them invoke other (already synthesized) programs.
Consider for example the problem of writing a program for a ticket-vendor application that can, among other things, issue and reserve tickets.
Building such a system requires creating modules for various data structures---perhaps a stack and queue---and using these modules in a top-level module that processes ticket requests. 
It is natural to ask whether such modules can be synthesized separately---i.e., in a compositional fashion.

% From the perspective of what works in program synthesis---and what does not work---it is clear that some kind of compositional approach is vital.
% 
The fundamental question is
\begin{tcolorbox}[boxsep=0pt,left=3pt,right=3pt,top=3pt,bottom=3pt]
  Can one address the scalability problem of program synthesis by exploiting compositionality, so that (i) larger programs are synthesized from smaller programs, and (ii) each ``larger-created-from-smaller'' synthesis sub-problem is of a similar nature, so that the essence of each sub-problem (and the size of each search space) has roughly the same character?
\end{tcolorbox}

\noindent
% While the principle of compositionality has been successfully exploited in many other domains---e.g., program verification---the synthesis community has not identified the means for applying this principle to program synthesis.
A solution to this question is surprisingly tricky to envisage. 
Most existing synthesis approaches require having a concrete semantics or implementation in hand when reasoning about modules, components, APIs, etc.~\cite{DBLP:conf/popl/FengM0DR17,DBLP:journals/pacmpl/ShiSL19,singh14}, and such synthesis tools end up reasoning about the entire program all the way down to its lowest-level components.
Not only is this approach in fundamental opposition to the ``similar-nature/similar-size'' principle articulated above, it makes synthesis increasingly hard as more modules are considered.
% 
% Returning to our earlier example, existing approaches require reasoning about concrete implementations of the queue and stack modules when synthesizing the ticket-vending module, thus making synthesis increasingly harder as more modules are considered.

Instead, when code is synthesized for some module $M$, all reasoning about lower-level modules $\{ M_i \}$ on which $M$ \emph{directly} depends should be carried out in a way that is \emph{agnostic} about the implementations of $\{ M_i \}$.
% The inability to perform implementation-agnostic reasoning is a key shortcoming of existing synthesis approaches.
This observation leads us to pose two related challenges:
\rone  
  How can one carry out program synthesis without having in hand details about the implementations of lower-level modules?
\rtwo
  How can one ensure that each synthesis problem results in code that is independent of the implementations of lower-level modules?

% I: Innovation
In this paper, we present the case for the following thesis:
\begin{tcolorbox}[boxsep=0pt,left=3pt,right=3pt,top=3pt,bottom=3pt]
Program synthesis can scale using modular system design.
\end{tcolorbox}

\noindent
Modular system design is one of the most important concepts in designing software.
A system should be organized in a layered fashion, where information hiding is used to hide implementation choices \cite{CACM:Parnas72}.
The \textit{information-hiding} principle intuitively states that each module exports an interface that does not reveal specific implementation choices used inside the module, and changing the module's implementation should not force any changes to be made to other modules.
% 
% This principle enables large-scale software development through compositionality---large programs are created from smaller programs---however, information hiding goes beyond compositionality \emph{per se}.
% In particular, by imposing restrictions on what operations are visible, it promotes \emph{malleability}: a system that respects information hiding has a greater capacity for adaptive change---e.g., one can change the implementation of the queue module without worrying about changes needing to be made to the client ticket-vending module.

Programmers practice modular system design, or at least aspire to it.
In essence, our goal is to  provide a level of automation for what good programmers do manually.
Of course, we are not trying to automate everything.
What is left in the hands of the programmer are architectural decisions
and specifications of the intended behavior of individual modules.
The programmer is responsible for the overall organization of the system's design, and must decide such issues as:
What are the layers in the system?
What are the implementation choices in a given layer (such as choices about data structures and data representations)?
What operations are exposed in each layer, and what is the intended behavior of each operation?

We identify \textit{two} opportunities for providing automation for each module and, as a key contribution of this paper, we formally define these synthesis problems.

\noindent\textbf{Module-Implementation Synthesis.}
Synthesis can be helpful in creating the implementations of the various functions in each module from some specifications. 
The key difference from traditional synthesis problems is that implementation details of ``lower'' modules are not available.
Instead, one only has access to \emph{implementation-agnostic specifications} of the semantics of such modules.
% 
% For example, in our ticket-vending application, one might want to synthesize the functions that process tickets using a queue. 
% % 
% However, the queue module may only expose algebraic properties~\cite{DBLP:journals/acta/SpitzenW74,Thesis:Guttag75,CGPRDS:GTWW75,DBLP:journals/tse/LiskovZ75,DBLP:journals/acta/GuttagH78} that characterize the semantics of queue functions (e.g., $\ias{\modqueue}=$``\texttt{dequeue(enqueue(q,i)) = if (is\_empty(q)) then q else enqueue(dequeue(q),i)}''), while hiding internal details that are implementation-specific.

\noindent\textbf{Module-Specification Synthesis.}
Because modules can only expose their semantics to other modules in a way that does not reveal their implementation details, it can be challenging to come up with such semantic definitions.
We propose to automate the creation of such implementation-agnostic semantic definitions using synthesis, namely, \emph{synthesis of formulas}.
% 
% In our example, once one has synthesized an implementation of the queue module (via the previous type of synthesis problem), the system will automatically synthesize its implementation-agnostic semantic properties, such as $\ias{\modqueue}$, which establishes a relationship between enqueue and dequeue.

Note the role of the second kind of synthesis problem:
its results provide part of the specification when one moves on to the task of synthesizing the implementation of functions in the next module.
By analogy with the Paul Simon lyric ``one man's ceiling is another man's floor''
\cite{Simon:CeilingFloor73}, we have ``one module's semantics is another module's primitives.''

We call this approach \emph{modular system synthesis} (\framework).
The visibility restrictions of information hiding provide the key for \framework to achieve the objective of making synthesis scalable via ``similar-nature/similar-size'' sub-problems:
both of our synthesis problems concern a single module of the system, and a single module's implementation only. 
%
% \twr{The phrase ``by avoiding the use of explicit components'' makes it sound like we abandon modularity;
% it will confuse the reader.}
By concealing the implementation of lower-level modules, \framework ensures that the formula representing the semantics of these layers remains independent of the size of the ``accumulated'' system as we move to higher-level layers. 
Moreover, \framework retains the usual benefit of modular system design, namely, it results in software that (usually) can be readily adapted---in this context, re-synthesized---as requirements change.

This paper contributes both a framework and solidifying the concept of contract-based design in the context of program synthesis, which abstracts components or sub-systems based on their interfaces. Notably, the study of interface compatibility and composition has not been extensively explored in the context of program synthesis, opening up many opportunities for future developments.
Specifically, using the aforementioned ticket-vending application as an example~(\S\ref{Se:IllustrativeExample}), it
(i) defines modular system synthesis (\S\ref{Se:ModularSystemDesign});
(ii) defines the two kinds of synthesis problems
that arise in \framework (\S\ref{Se:SynthesisForModularSystemDesign}); and
(iii) describes a proof-of-concept system, called \tool, that achieves these goals (\S\ref{Se:ImplementationAndCaseStudy}).

\tool is based on two existing program-synthesis techniques: \jlibsketch~\cite{DBLP:journals/pacmpl/MarianoRXNQFS19} a program-sketching tool that supports algebraic specifications, and \spyro~\cite{https://doi.org/10.48550/arxiv.2301.11117} a tool for synthesizing precise specifications from a given codebase.
We used \tool to carry out case studies based on two-layer modular synthesis problems from Mariano et al.~\cite{DBLP:journals/pacmpl/MarianoRXNQFS19},
which demonstrated that concealing lower-level components can be advantageous in reducing the complexity of the synthesis problem. 
Expanding upon their work, our case study in \S\ref{Se:CaseStudyEvaluation} further explored scenarios involving multiple layers.
\framework exhibits even better scalability compared to scenarios where executable semantics for all lower layers are exposed.
A further case study based on Mariano et al.\ in \S\ref{Se:AdditionalCaseStudyEvaluationMariano} also highlights the challenges of writing correct specifications.
Our framework and the act of performing synthesis for both the implementations and specifications of the modules unveiled bugs in the modules synthesized by Mariano et al.\ and in the module's specifications, which they manually wrote.

\S\ref{Se:RelatedWork} discusses related work.
\S\ref{Se:Conclusion} concludes.

% \lorischanged{
% \subsection*{Final Checklist}
% \begin{itemize}
%   \item
%     Notation (and use of macros) for implementation-specific specifications and implementation-agnostic specification
%   \item
%     Eliminate name clashes in operation names at different levels:
%     \begin{itemize}
%       \item \texttt{List::empty}, \texttt{Stack::empty}, and \texttt{Queue::empty}
%       \item \texttt{List::is\_empty}, \texttt{Stack::is\_empty}, and \texttt{Queue::is\_empty}
%       \item \texttt{List::size}, \texttt{Stack::size}, and \texttt{Queue::size}
%     \end{itemize}
%   \item
%     Check references to figures
%     \item vender vs vendor
%     \item \textit{Queue} vs \texttt{Queue} etc in figures/defs/examples
% \end{itemize}
% }

%% file: 2-illustrative-example.tex
\section{Illustrative Example}
\label{Se:IllustrativeExample}

\noindent
We present an experiment that illustrates the various aspects of \framework.
The problem to be solved is as follows:
  Synthesize a simple ticket-vendor application that supports the operations $\preparesales$, $\reserveticket$, $\issueticket$, $\soldout$, $\numticket$, and $\numwaiting$.
    (To simplify matters, we assume it is not necessary to cancel a reservation.)

\subsection{A Modular \modticket Implementation}
We decompose the system into three modules (Fig.~\ref{fig:ticket-vendor-example}):

\noindent
\textbf{Module 3:}
The \modticket module uses a \modqueue of reservations to implement the aforementioned operations.
% $\preparesales$, $\reserveticket$, $\issueticket$, $\soldout$, $\numticket$, and $\numwaiting$.
    
\noindent
\textbf{Module 2:}
The \modqueue module implements the operations $\emptyqueue$, $\enq$, $\front$, $\deq$, $\qsize$, and $\isemptyqueue$.
In our setting, a \modqueue is implemented using two stacks \cite{DBLP:journals/ipl/HoodM81}.\footnote{
  The invariant is that the second \modstack holds a prefix of the \modqueue's front elements, with the top element of the second \modstack being the \modqueue's front-most element. The first \modstack holds the \modqueue's back elements---with the top element of the first \modstack being the \modqueue's back-most element.
}
    
\noindent
\textbf{Module 1:}
The \modstack module implements the operations $\emptystack$, $\push$, $\sttop$, $\pop$, $\stsize$, and $\isemptystack$. 
In our setting, a \modstack is implemented using linked-list primitives of the programming language.

Moreover, the implementation of each module is to abide by the principle of information hiding:
\rone The \modticket module can use  operations exposed by \modqueue, but their actual implementations are hidden in Module 2.
\rtwo
  The \modqueue module can use  operations exposed by \modstack, but their actual implementations are hidden in Module 1.

\subsection{The Input of Modular \modticket Synthesis}

A \tool user supplies the following information:

\vspace{2pt}
\noindent
\textbf{\textit{Architectural-design choices}}:
\begin{itemize}
      \item
        The decomposition of the problem into \modticket, \modqueue, and \modstack modules (gray boxes in Fig.~\ref{fig:ticket-vendor-example}).
      \item
        Which operations are to be exposed by each module, denoted by $\PL[module]$---e.g., in Fig.~\ref{fig:ticket-vendor-example}, the \modqueue module exposes $\PL[\modqueue]$, which contains $\enq$ and $\deq$ operations, but not $\push$ and $\pop$ operations on the underlying stacks.
\end{itemize}

\noindent
\textbf{\textit{Data-structure/data-representation choices}}:

\noindent
\textbf{Module 3:}
        \modticket uses a \modqueue.

\noindent
\textbf{Module 2:}
        A \modqueue is implemented using two {\modstack}s.

\noindent
\textbf{Module 1:}
        A \modstack is implemented using a linked list.

\noindent
These choices are shown by the green boxes underneath each module in Fig.~\ref{fig:ticket-vendor-example}. 
For example, the \modqueue module is built on top of the \modstack module. 
However, only the \modstack interface---i.e., the function symbols in $\PL[\modstack]$ and its (potentially synthesized) implementation-agnostic specification $\ias{\modstack}$---is accessible by the \modqueue module.

\vspace{2pt}
\noindent
\textbf{\textit{Specifications of the module-specific synthesis problems}}:

\noindent
\textbf{Module 3:}
        Specifications of the behaviors of $\preparesales$, $\reserveticket$, $\issueticket$, $\soldout$, $\numticket$, and $\numwaiting$ in terms of the exposed \modqueue operations (and possibly other \modticket operations).
        For example, the implementation-specific specifications for the \modticket module, denoted by the yellow box labeled $\iss{\modticket}$ in Fig.~\ref{fig:ticket-vendor-example}, might constrain $\issueticket$ to dequeue a buyer from the underlying \modqueue module, but only if $\soldout$ (a \modticket operation) is false.

      \noindent
\textbf{Module 2:}
        Specifications of the behaviors of the \modqueue operations in terms of the exposed \modstack operations (and possibly other \modqueue operations).
        For example, the implementation-specific specification for the \modqueue module ($\iss{\modqueue}$), shown in Fig.~\ref{fig:ticket-vendor-example}, contains, among others, constraints that state that
        \rone if the first stack $st_{in}$ is empty, so is the second stack $st_{out}$,
        \rtwo enqueuing 1 on an empty queue and then retrieving the front of the queue yields 1.  
        
\noindent
\textbf{Module 1:}
        Specifications of the behaviors of the \modstack operations in terms of the programming language's linked-list operations (and possibly other \modstack operations). 
        For example, the implementation-specific specification of the \modstack module ($\iss{\modstack}$) might specify that $\push$ adds an element on the front of the stack's underlying linked list.

A user must also specify a search space of possible implementations. 
In \tool, this is done using a \sketch file.

%%%%
%% pic (subelement) definitions
%%%%
\tikzset{
    iasrect/.pic={
        \filldraw[dashed, very thick,fill=green!10,text=black] (-0.5,0) rectangle (1.6,0.9) node[midway] {$\ias{#1}$};
    },
    % iasrectdouble/.pic={ % This exists for the base language ias, which is user-given
    %     \filldraw[thick,fill=green!10,text=black] (0,0) rectangle (2.2,1) node[midway] {$\ias{#1}$};
    % },
    issrect/.pic={
        \filldraw[thick,fill=yellow!10,text=black] (-0.5,0) rectangle (5.5,1) node[midway] {$\iss{#1}$};
    },
    plrectout/.pic={
        \filldraw[fill=white!10,fill opacity = 1,text=black] (-0.5,0) rectangle (5.5,1) { };
    },
    plrect/.pic={
        \filldraw[thick,fill=green!10,text=black] (-0.5,0) rectangle (1.6,0.9) node[midway] {$\PL[#1]$};
    },
    implrect/.pic={
        \filldraw[dashed,very thick,fill=blue!10,text=black] (-0.5,0) rectangle (5.5,1) node[midway] {#1 Implementation};
    },
    modbox/.pic={
        \filldraw[fill=black!6] (0,0) rectangle (7,4);
        \node[anchor=north west] at (0,3.95) {#1};
    },
    modulebase/.pic={
        \draw (0,0) pic {modbox=#1};        
        \draw (1,4) pic {plrectout=#1};
        \draw (1.2,4) pic {plrect=#1};
        \draw (4.7,4) pic {iasrect=#1};
        \draw (1,0) pic {issrect=#1};
        \draw (1,2) pic {implrect=#1};
        \draw[->,very thick] (5.25,1) -- (5.25,2) node[midway,anchor=east] {Implementation Synthesis};
        \draw[->,very thick] (5.25,3) -- (5.25,4) node[midway,anchor=east] {Specification Synthesis};
        \draw[->,very thick] (1.75,-1) -- (1.75,0);
        \draw[->,very thick] (5.25,-1) -- (5.25,0);
        % \draw[dashed] (3.5,5) -- (4.5,5);
    },
    moduleimpl/.pic={
        \draw (1,0) pic {issrect=#1};
        \draw (1,2) pic {implrect=#1};
    },
    module/.pic={ % Size: 5
        \draw (0,0) pic {modulebase=#1};
        \draw (0,0) pic {moduleimpl=#1};
    },
    basepl/.pic={
        \draw (1,0) pic {plrectout=\modlist};
        \draw (4.7,0) pic {iasrect=\modlist};
        \draw (1.2,0) pic {plrect=\modlist};
    },
    ticketvendor/.pic={
        \draw (0,0) pic {modbox=\modticket};
        \draw (1,0) pic {issrect=\modticket};
        \draw (1,2) pic {implrect=\modticket};
        \draw[->,very thick] (5.25,1) -- (5.25,2) node[midway,anchor=east] {Implementation Synthesis};
        \draw[->,very thick] (1.75,-1) -- (1.75,0);
        \draw[->,very thick] (5.25,-1) -- (5.25,0);
    },
    stackplcallout/.pic={
        \fill[fill=green!10,text=black] (0,-0.8) rectangle (6.25,4);
        \draw[very thick] (0,3) -- (0,4);
        \draw[very thick] (0,4) -- (6.25,4);
        \draw[very thick] (6.25,3) -- (6.25,4);        
        \draw[dashed,very thick] (0,-0.8) -- (0,3);
        \draw[dashed,very thick] (0,-0.8) -- (6.25,-0.8);
        \draw[dashed,very thick] (6.25,-0.8) -- (6.25,3);
        \draw (0,4) node[anchor=north west,align=left,text width=6.3cm]
        {
        $\textbf{Functions exposed in }\mathbf{\PL[\modstack]}$:
        \\ 
        $\emptystack$, $\push$, $\pop$, $\sttop$, $\stsize$, $\isemptystack$
        \\\;\\ % a blank line
        $\textbf{Implementation-agnostic spec }\mathbf{\ias{\modstack}}$:
        \\
        $\begin{array}{lr@{\hspace{0em}}}
        \isemptystack(\emptystack)  =  \tru \\
        \isemptystack(\push(st,x)) = \fls\\
        \sttop(\push(st,x)) = x & \callout{1}\\              
        \pop(\push(st,x)) = x\\
        \stsize(\emptystack) = 0\\
        \multicolumn{2}{l}{\stsize(\push(st,x)) = \stsize(st) + 1}\\
        \end{array}$
        };
    },
    queueplcallout/.pic={
        \fill[fill=green!10,text=black] (0,0.2) rectangle (6.25,5.5);
        \draw[very thick] (0,4.2) -- (0,5.5);
        \draw[very thick] (0,5.5) -- (6.25,5.5);
        \draw[very thick] (6.25,4.2) -- (6.25,5.5);        
        \draw[dashed,very thick] (0,0.2) -- (0,4.2);
        \draw[dashed,very thick] (0,0.2) -- (6.25,0.2);
        \draw[dashed,very thick] (6.25,0.2) -- (6.25,4.2);
        \draw (0,5.5) node[anchor=north west,align=left,text width=6.3cm]
        {        
        $\textbf{Functions exposed in }\mathbf{\PL[\modqueue]}$:
        \\ 
        $\emptyqueue$, $\enq$, $\deq$, $\front$, $\isemptyqueue$, $\qsize$
        \\\;\\ % a blank line
        $\textbf{Implementation-agnostic spec }\mathbf{\ias{\modqueue}}$:\\
        $\begin{array}{l@{\hspace{0.5em}}l@{\hspace{0em}}r@{\hspace{0em}}}
            \multicolumn{2}{l}{\front(\enq(q, x)) =} & \\
            & \ite(\isemptyqueue(q), x, \front(q)) &\\
            \multicolumn{2}{l}{\deq(\enq(q, x)) =} & \callout{4} \\
            & \ite(\isemptyqueue(q), \emptyqueue, &\\
            & \qquad\enq(\deq(q), x))\\
            \multicolumn{2}{l}{\qsize(\enq(q, x)) = \qsize(q) + 1} & \\
            \multicolumn{2}{l}{\ldots} &\\
        \end{array}$
        };
    },
    queueisscallout/.pic={
        \filldraw[thick,fill=yellow!10,text=black] (0,-0.3) rectangle (6.25,2.5);
        \draw (0,2.5) node[anchor=north west,align=left,text width=6.3cm]
        {         
        $\textbf{Implementation-specific spec. }\mathbf{\iss{\modqueue}}$:\\
        $\begin{array}{l@{\hspace{0em}}r@{\hspace{-0.2em}}}
        \multicolumn{2}{l}{\isemptystack(st_{out}) \rightarrow \isemptystack(st_{in})} \\
        \front(\enq(\emptyqueue, 1)) = 1 & \\
        \isemptyqueue(\enq(\emptyqueue, 3)) = \fls & \callout{2}\\
        \qsize(\enq(\emptyqueue, x)) = 1 & \\
        \ldots & \\
        \end{array}$
        };
    },
    queueimplcallout/.pic={
        \filldraw[dashed,very thick,fill=blue!10,text=black] (0,-0.3) rectangle (6.25,3);
        \draw (0,3) node[anchor=north west,align=left,text width=6.3cm]
        {
        \textbf{\modqueue Implementation}:\\
        \modqueue = ($st_{in}$: \modstack, $st_{out}$: \modstack) \\
        \vspace{1ex}
        $\begin{array}{ll@{\hspace{0em}}r@{\hspace{-0.3em}}}
        \multicolumn{3}{l}{\enq(q: \modqueue, i: \inttype): \modqueue = } \\
        & \texttt{if } \isemptystack(q.st_{out}) & \callout{3} \\
        & \multicolumn{2}{l}{\texttt{then } (q.st_{in}, \push(q.st_{out}, i))} \\
        & \multicolumn{2}{l}{\texttt{else } (\push(q.st_{in}, i), q.st_{out})} \\
        \ldots
        \end{array}$
        };
    },
    queuelistisscallout/.pic={
        \filldraw[thick,fill=yellow!10,text=black] (0,-0.3) rectangle (6.25,3);
        \draw (0,3) node[anchor=north west,align=left,text width=6.3cm]
        {
        $\textbf{Implem.-specific spec. }\mathbf{\iss{\modqueue (as~\modlist)}}$:\\
        $\begin{array}{lr}
        \isemptylist(\emptyqueue.l) \\
        % $\enq(\emptyqueue, 1).l = \lcons(1, \lnil)$ \\
        % $\enq(\enq(\emptyqueue, 1), 2).l =$ \\ 
        % \qquad $\lcons(1, \lcons(2, \lnil))$ \\ 
        \front(q) = \lhead(q.l) \\
        \front(\enq(\emptyqueue, 1)) = 1 & \callout{5} \\
        \isemptyqueue(\enq(\emptyqueue, 3)) = \fls \\
        \qsize(\enq(\emptyqueue, x)) = 1 \\
        \ldots 
        \end{array}$
        };
    },
    queuelistimplcallout/.pic={
        \filldraw[dashed,very thick,fill=blue!10,text=black] (0,0.5) rectangle (6.25,3);
        \draw (0,3) node[anchor=north west,align=left,text width=6.3cm]
        {
        \textbf{\modqueue (as~\modlist) Implementation}:\\
        % $\mathbf{\iss{\modqueue (as~\modlist)}}$:\\
        \modqueue = ($l$: \modlist) \\
        \vspace{1ex}
        $\enq(q: \modqueue, i: \inttype): \modqueue =$ \\ 
        \qquad $(\lsnoc(q.l, i))$\\
        \ldots
        };
    },
    labelpoint/.pic={
        \node[anchor=south west] at (0,0) {(#1)};
    }
}

%%%
% Main example figure
%%%
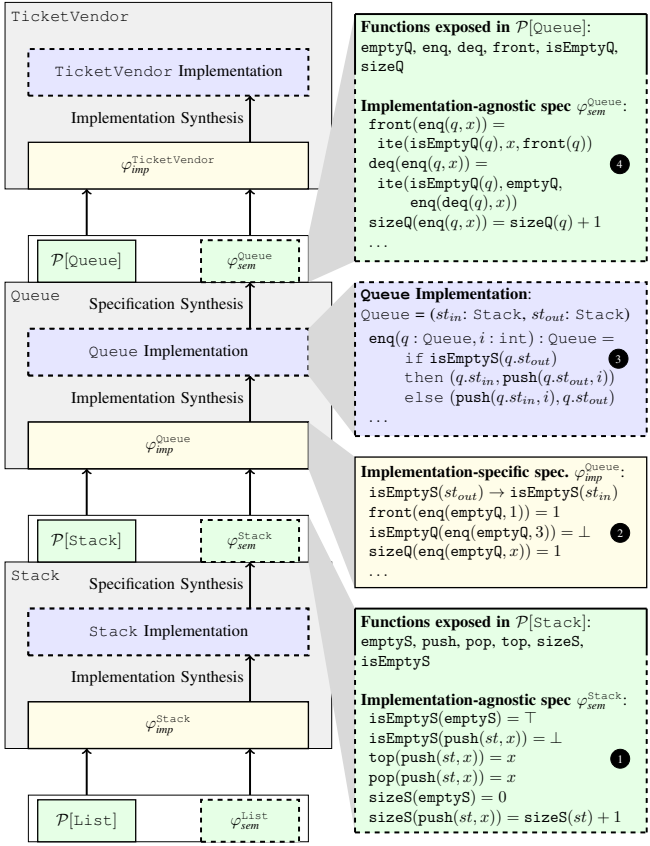
\begin{figure}[tb!]
    \centering
    \resizebox{0.489\textwidth}{!}{
    \begin{tikzpicture}[]
        % Module stack
        \draw (0,6) pic {ticketvendor};
        \draw (0,0) pic {module=\modqueue};
        \draw (0,-6) pic {module=\modstack};
        \draw (0,-8) pic {basepl};

        % \modqueue PL/phisem callout
        \fill[fill=gray!30] (7.5,4.45) -- (6.5,4) -- (6.5,5) -- (7.5,9.75) -- cycle;
        \draw (7.5,4.25) pic {queueplcallout};
        %\draw (9,4.5) -- (7,4);
        %\draw (9,10) -- (7,5);

        % \modqueue impl callout
        \fill[fill=gray!30] (7.5,0.7) -- (6.5,2) -- (6.5,3) -- (7.5,4) -- cycle;
        \draw (7.5,1) pic {queueimplcallout};
        %\draw (9,0.5) -- (7,2);
        %\draw (9,3.5) -- (7,3);

         % \modqueue iss callout
        \fill[fill=gray!30] (7.5,-2.57) -- (6.5,0) -- (6.5,1) -- (7.5,0.25) -- cycle;
        \draw (7.5,-2.25) pic {queueisscallout};
        %\draw (9,-3) -- (7,0);
        %\draw (9,-0.5) -- (7,1);

        % Stack PL/phisem callout
        \fill[fill=gray!30] (7.5,-7.8) -- (6.5,-2) -- (6.5,-1) -- (7.5,-3) -- cycle;
        \draw (7.5,-7) pic {stackplcallout};        
        %\draw (9,-8) -- (7,-2);
        %\draw (9,-4) -- (7,-1);

        % Reference points - seems unnecessary - everything we refer to has a label
        %\draw (0,10) pic {labelpoint=a};
        %\draw (0,4) pic {labelpoint=b};
        %\draw (0,-2) pic {labelpoint=c};
    \end{tikzpicture}
    }
    \caption{
    Organization of the modular \modticket synthesis problem: user-supplied inputs are shown in solid boxes;
    synthesized outputs are shown in dashed boxes. On the right, the \modqueue module's specifications and implementation are expanded; the other modules would have similar details.
    }
    \vspace{-3.0ex}
    \label{fig:ticket-vendor-example}
\end{figure}

\subsection{The Output of Modular \modticket Synthesis}
\label{Se:TheOutputOfModularTicketVendorSynthesis}

Using the \framework framework, we synthesize three module implementations:
the \modticket module implementation, which satisfies $\iss{\modticket}$ (and uses \modqueue);
the \modqueue module implementation, which satisfies $\iss{\modqueue}$ (and uses \modstack);
and the \modstack module implementation, which satisfies $\iss{\modstack}$ (and uses lists).
However, to  synthesize the \modticket module implementation, we need an \textit{implementation-agnostic specification} of \modqueue, denoted by $\ias{\modqueue}$.
The same can be said for the \modqueue module implementation, for which we need an implementation-agnostic specification of \modstack, denoted by $\ias{\modstack}$.\footnote{
  Technically, \modlist is part of the programming language;
  however, so that all sub-problems have the same form, we assume---as shown in Fig.~\ref{fig:ticket-vendor-example}---that we also have available an implementation-agnostic specification of \modlist, denoted by $\ias{\modlist}$.
  In our evaluation, we synthesize $\ias{\modlist}$ automatically.
}

The user could write $\ias{\modqueue}$ and $\ias{\modstack}$ manually, but it is more convenient to synthesize these specifications from the \modqueue and \modstack module implementations, respectively.
The \framework methodology is to start with the bottom-most module and work upward, alternately applying two synthesis procedures:
first synthesizing the implementation of a module $M$ and then synthesizing $M$'s implementation-agnostic specification $\ias{M}$, which gets exposed to the next higher module.

For the modular \modticket-synthesis problem, we start with \modstack, the bottommost module, and synthesize a \modstack module implementation---a set of $\PL[\modlist]$ programs---that satisfies the implementation-specific specification $\iss{\modstack}$. 
(In \tool, this step is done using program sketching and the tool \jlibsketch~\cite{DBLP:journals/pacmpl/MarianoRXNQFS19}.)
This step is depicted in Fig.~\ref{fig:ticket-vendor-example} as the {Implementation Synthesis} problem in the \modstack module. 
We then switch to the {Specification Synthesis} problem for \modstack, and synthesize $\ias{\modstack}$, an implementation-agnostic specification of \modstack.
(In \tool, this step is done by providing a grammar of possible properties and by using the tool \spyro~\cite{https://doi.org/10.48550/arxiv.2301.11117}.)
For the \modstack module, the resultant $\ias{\modstack}$ is the conjunction of the  equalities shown at \callout{1} in Fig.~\ref{fig:ticket-vendor-example}.
% \begin{equation}
%   \begin{array}{@{\hspace{0ex}}r@{\hspace{0.5ex}}c@{\hspace{0.5ex}}l@{\hspace{2.5ex}}r@{\hspace{0.5ex}}c@{\hspace{0.5ex}}l@{\hspace{0ex}}}
%     \isemptystack(\emptystack) & = & \tru      & \sttop(\push(st,x)) & = & x \\
%     \isemptystack(\push(st,x)) & = & \fls      & \pop(\push(st,x)) & = & st \\
%     \stsize(\emptystack) & = & 0  & & & \\
%     \multicolumn{6}{c}{\stsize(\push(st,x))  = \stsize(st) + 1}
%   \end{array}
%   \label{Eq:stack-impl-ag-spec}
% \end{equation}
% This step is shown in Fig.\,\ref{fig:ticket-vendor-example} as the Specification Synthesis step.

Using $\ias{\modstack}$ (\callout{1}), together with the implementation-specific specification $\iss{\modqueue}$ (\callout{2}), we now synthesize the \modqueue module implementation (\callout{3})---a set of $\PL[\modstack]$ programs---and the implementation-agnostic specification $\ias{\modqueue}$ (\callout{4}) via the same two-step process.

Finally, using $\ias{\modqueue}$ and the implementation-specific specification $\iss{\modticket}$, we synthesize the \modticket module implementation.
(If needed by a further client, we would then synthesize the implementation-agnostic specification $\ias{\modticket}$.)
Thus, the last output of the synthesis procedure, shown in Fig.~\ref{fig:ticket-vendor-example}, consists of implementations of \modstack, \modqueue, and \modticket, and the implementation-agnostic specifications $\ias{\modstack}$ and $\ias{\modqueue}$.
% (Discussion of the synthesized implementations of \modstack, \modqueue, and \modticket is deferred until \S\ref{Se:CaseStudyEvaluation}.)
% (The synthesized implementations of \modstack, \modqueue, and \modticket are discussed in \S\ref{Se:CaseStudyEvaluation}.)

%%%
% Alternative \modqueue Implementation Figure
%%%
\begin{figure}[tb!]
    \centering
    \resizebox{0.489\textwidth}{!}{
    \begin{tikzpicture}[]
        % Base and impl separately, as they have different names
        \draw (0,0) pic {modulebase=\modqueue};
        \draw (0,0) pic {moduleimpl=\modqueue (as~\modlist)};
        \draw (0,-2) pic {basepl};

        % Implementation callouts
        \fill[fill=gray!35] (7.5,4.5) -- (6.5,3) -- (6.5,2) -- (7.5,2) -- cycle;
        \draw (7.5,1.5) pic {queuelistimplcallout};
        \fill[fill=gray!35] (7.5,1.51) -- (6.5,1) -- (6.5,0) -- (7.5,-1.82) -- cycle;
        \draw (7.5,-1.5) pic {queuelistisscallout};

        % Callout lines
        %\draw[thin] (9,4.5) -- (7,3);
        %\draw[thin] (9,1.5) -- (7,2);
        
        %\draw[thin] (9,1) -- (7,1);
        %\draw[thin] (9,-2) -- (7,0);
        
    \end{tikzpicture}
    }
    \caption{Alternative implementation of the \modqueue module using list primitives instead of two stacks. $\PL[\modqueue]$ and $\ias{\modqueue}$ are the same as in Fig. \ref{fig:ticket-vendor-example}.}
    \label{fig:ticket-vendor-example/queuelist}
    \vspace{-3.0ex}
\end{figure}

\subsection{Benefits of Modular System Synthesis}
\label{Se:DiffQueue}
%% This is a good place to point out one of the key advantages of using information-hiding.
% 
At some point, we might want to decide to modify the implementation of the \modqueue module to use directly the linked-list primitives provided by the language (shown in Fig.~\ref{fig:ticket-vendor-example/queuelist}).
Information hiding allows us to do so in a compartmentalized way---i.e., by only changing the specific \modqueue module.
Importantly, the module's interface, composed of the function symbols in $\PL[\modqueue]$ and its implementation-agnostic specification $\ias{\modqueue}$,
does not change when the implementation of the \modqueue module changes.
Because this interface is what the \modticket module was synthesized with respect to, changes to the \modqueue implementation are not visible to \modticket.

%% file: 3-modular-system-design.tex
\section{Modular System Design}
\label{Se:ModularSystemDesign}

In this section, we formally define modular system design and the corresponding specification mechanisms.
A system is organized in modules, and each module exports a module interface $\MI$ and a specification $\ias{\MI}$ of the semantics of the module interface.
Both $\MI$ and $\ias{\MI}$ hide the module's implementation.
A module's implementation can also have a set of private functions $\PF$,  which can only be used within the module.
A program is constructed by stacking layers of such modules.\footnote{
  In general, the structure of the dependencies among layers can form a directed acyclic graph.
  However, to reduce notational clutter, throughout the paper we assume that the layers have a strict linear order.
  % The generalization to dag-structured layers is straightforward.
}
For instance, the example in Fig.~\ref{fig:ticket-vendor-example} has three modules: \modstack, \modqueue, and \modticket.
(None of those modules have private functions.)

In the following, we assume a programming language $\PL$ (e.g., C with its core libraries),
and use $\PL[\MI]$ to denote $\PL$ extended with the functions exposed by module $\MI$.
% To simplify matters, we consider calls to functions in the APIs of core libraries, such as \texttt{glibc} for GNU/Linux systems, to be operations intrinsic to $\PL$.

\begin{definition}[Modular System Design]\label{De:ModularSystemDesign}
A system is \textbf{\emph{implemented modularly}} if it is partitioned into disjoint sets of functions $\PF_1, \MI_1, \PF_2, \MI_2, \ldots, \PF_n, \MI_n$, such that for each $f \in \PF_i \cup \MI_i$, $f$ is implemented using $\PL[\MI_{i-1} \cup \PF_i \cup \MI_i]$---i.e., $f$ only uses operations in $\PL$, 
and calls to functions in the interface exported from layer $i\textit{--}1$, to private functions of layer $i$, and to functions in the interface exported from layer $i$.
\end{definition}

To reduce notational clutter, we will ignore private functions, and only discuss the functions in module interfaces.

As we saw in \S\ref{Se:IllustrativeExample}, we need to abide by the principle of \textit{information hiding}---i.e.,
changing the implementations of any function in $\MI_{i-1}$ should not require changing the implementations of functions in $\MI_i$.
With this principle in mind, we now describe the different natures of the specification for the module implementation at a given layer $i$ (\S\ref{Se:ISS}) and the specification exposed to layer $i+1$ (\S\ref{Se:IAS}).

\subsection{Implementation-specific Specifications}
\label{Se:ISS}

When synthesizing specific implementations of the functions $\MI_i$ at layer $i$, the specifications are allowed to use symbols in $\PL[\MI_{i-1} \cup \MI_i]$---i.e., the specification can refer to the functions we are specifying
and to the ones in the interface exported from the previous layer---as well as implementation-specific details from layer $i$ (e.g., data-structure declarations).

\begin{definition}\label{De:ISS}
An \textbf{implementation-specific specification} for a set of functions $\MI_i$ at layer $i$ 
% in a system implemented in a modular fashion 
is a predicate $\iss{{\MI_i}}$ that only uses symbols in $\PL[\MI_{i-1}\cup \MI_{i}]$.
\end{definition}

\begin{example}
In the implementation-specific specification of \modqueue from Fig.\ \ref{fig:ticket-vendor-example}, where \modqueue is implemented using two {\modstack}s, one of the properties is as follows:
\[
  \isemptyqueue(q) \iff \isemptystack(q.st_{in}) \wedge \isemptystack(q.st_{out}).
\]
For the version from Fig.\ \ref{fig:ticket-vendor-example/queuelist}, where \modqueue is implemented using a \modlist, the analogous property is
\[
  \isemptyqueue(q) \iff \mathtt{\isemptylist}(q.l).
\]

A specification might also contain a set of examples, e.g., $\front(\enq(\emptyqueue, 1)) = 1$ and $\front(\enq(\enq(\emptyqueue, 1), 2)) = 1$.
\end{example}

\subsection{Implementation-agnostic Specifications} 
\label{Se:IAS}

While implementation-specific details are needed to converge on an implementation with which the programmer is happy, when exposing the specification of $\MI_i$ at layer $i+1$, to abide to the principle of information hiding, one cannot provide specifications that involve function symbols in $\PL[\MI_{i-1}\cup \MI_{i}]$, but only those in $\PL[\MI_{i}]$.

\begin{definition}\label{De:IAS}
An \textbf{implementation-agnostic specification} for a set of functions $\MI_i$ at layer $i$ is a predicate
$\ias{{\MI_i}}$ that only uses symbols in $\PL[\MI_{i}]$.
\end{definition}

\begin{example}
Because of the vocabulary restrictions imposed by Def.\ \ref{De:IAS}, it is natural for implementation-agnostic specifications to take the form of algebraic specifications
\cite{DBLP:journals/acta/SpitzenW74,Thesis:Guttag75,CGPRDS:GTWW75,DBLP:journals/tse/LiskovZ75,DBLP:journals/acta/GuttagH78}.
For instance, for the \modqueue module, the conjunction of the following equalities is an implementation-agnostic specification $\ias{\modqueue}$ for \modqueue:
{\small
\begin{equation}
  \label{Eq:QueueIAS}
  \hspace{-2mm}
  \begin{array}{l} 
    \isemptyqueue(\emptyqueue) = \tru \hspace{5pt}
    \isemptyqueue(\enq(q, x)) = \fls \\
    \qsize(\emptyqueue) = 0  \qquad~
    \qsize(\enq(q, x)) = \qsize(q) + 1 \\
    \front(\enq(q, x)) = \ite(\isemptyqueue(q), x, \front(q)) \\
    \deq(\enq(q, x)) = \ite(\isemptyqueue(q), q, \deq(\enq(q), x))\\
  \end{array}
\end{equation}
}

Note that Eq.\ \eqref{Eq:QueueIAS} serves as $\ias{\modqueue}$ both for the version of \modqueue from Fig.\ \ref{fig:ticket-vendor-example}, where \modqueue is implemented using two {\modstack}s, and for the version of \modqueue from Fig.\ \ref{fig:ticket-vendor-example/queuelist}, where \modqueue is implemented using a \modlist.
\end{example}

%% file: 4-synthesis.tex
\section{Synthesis in Modular System Synthesis}
\label{Se:SynthesisForModularSystemDesign}

In this section, we define the \emph{implementation-synthesis} (\S\ref{Se:SynthesisOfImplementations}) and \emph{specification-synthesis} (\S\ref{Se:SynthesisOfImplementationAgnosticSpecifications}) problems that enable our scheme for modular system synthesis.

\subsection{Synthesis of Implementations}
\label{Se:SynthesisOfImplementations}
The obvious place in which synthesis can be helpful is in synthesizing the implementations of the various functions at each layer from their implementation-specific specifications.
For example, in Fig.~\ref{fig:ticket-vendor-example}, an implementation of \modqueue (the function $\enq$ is shown in the second box on the right) is synthesized from the implementation-agnostic specification 
$\ias{\modstack}$ of \modstack, \emph{and} an implementation-specific specification $\iss{\modqueue}$ that is allowed to talk about how the two {\modstack}s used to implement a \modqueue are manipulated (e.g., $\isemptystack(st_{out}) \rightarrow \isemptystack(st_{in})$).
    
\begin{definition}[Implementation synthesis]
For module interface $\MI_i$, the \textbf{implementation-synthesis} problem is a triple $(\searchspace_i, \ias{\MI_{i-1}}, \iss{\MI_i})$, where
\begin{itemize}
  \item
    $\searchspace_i$ is the set of possible implementations we can use for $\MI_{i}$ (every program in $\searchspace_i$ uses only symbols in $\PL[\MI_{i-1}\cup \MI_{i}]$).
    % (Typically, $S_i$ is given as a regular-tree grammar \cite{comon:hal-03367725} for a fragment of $\PL[\MI_{i-1}\cup \MI_{i}]$.)
  \item
    $\ias{\MI_{i-1}}$ is an implementation-agnostic specification of the module-interface functions in $\MI_{i-1}$.
  \item 
    $\iss{\MI_i}$ is an implementation-specific specification that uses only symbols in $\PL[\MI_{i-1}\cup \MI_{i}]$.
\end{itemize}
A solution to the implementation-synthesis problem is an implementation of $\MI_i$ in $\searchspace_i$ that satisfies $\iss{\MI_i}$.
\end{definition}

This particular form of synthesis where one draws a program from a search space to match a specification is fairly standard in the literature.
However, we observe that a particular aspect of modular system design makes most synthesis approaches inadequate---i.e., the specification $\ias{\MI_{i-1}}$ can talk about functions in $\MI_{i-1}$ only in an implementation-agnostic way.
For example, when synthesizing functions in \modqueue, we do not have direct access to a stack implementation---i.e., we cannot actually execute the implementation. 
Instead, we have access to the semantics of \modstack through implementation-agnostic properties such as $\isemptystack(\push(st, x)) = \fls$.

We are aware of only one tool, \jlibsketch, that can perform synthesis with algebraic specifications~\cite{DBLP:journals/pacmpl/MarianoRXNQFS19}, and we use it in our evaluation. 
In \jlibsketch, one provides $\searchspace_i$ as a program sketch (i.e., a program with integer holes that need to be synthesized), $\ias{\MI_{i-1}}$ as a set of rewrite rules over the functions in $\MI_{i-1}$, and $\iss{\MI_{i}}$ as a set of assertions.

\subsection{Synthesis of Implementation-agnostic Specifications}
\label{Se:SynthesisOfImplementationAgnosticSpecifications}

Because the implementation of layer $i\textit{-}1$ is hidden when performing synthesis at layer $i$, the user has to somehow come up with implementation-agnostic specifications like the ones shown in Fig.~\ref{fig:ticket-vendor-example}.
Our next observation is that such specifications can also be synthesized!
With this observation, modular system design becomes a fairly automatic business where the programmer mostly has to decide how to structure modules and provide implementation-specific specifications and search spaces (typically as regular-tree grammars \cite{comon:hal-03367725}).

In Fig.~\ref{fig:ticket-vendor-example}, the implementation-agnostic specification $\ias{\modqueue}$ of \modqueue is synthesized from the \modqueue implementation.
(The same $\ias{\modqueue}$, or one equivalent to it, is synthesized from the alternative \modqueue implementation of Fig.~\ref{fig:ticket-vendor-example/queuelist}.)

\begin{definition}[Specification synthesis]\label{De:IASSynthesis}
For module interface $\MI_i$, a \textbf{specification-synthesis problem} is a pair $(F_i,\propertyspace_i)$ where
\begin{itemize}
  \item
    $F_i$ is a set of programs, written in $\PL[\MI_{i-1}\cup \MI_i]$, that is a concrete implementation of $\MI_i$.
  \item
    $\propertyspace_i$ is the set of possible properties we can use for $\ias{\MI_i}$
    (every property in $\propertyspace_i$ uses only symbols in $\PL[\MI_i]$).
    (Typically, $\propertyspace_i$ is given as a regular-tree grammar for a fragment of logic in which terms can only use symbols in $\PL[\MI_i]$.)
\end{itemize}
A solution to the specification-synthesis problem is a set of properties $\ias{\MI_i} \subseteq \propertyspace_i$ such that for every $\iasprop\in \ias{\MI_i}$:
\begin{description}
    \item[Soundness:] The implementation $F_i$ satisfies $\iasprop$.
    \item[Precision:] There is no property $\iasprop'\in \propertyspace_i$ that implies $\iasprop$ and such that the implementation $F_i$ satisfies $\iasprop'$.
\end{description}
\end{definition}

In general, there might not be just one answer to this synthesis problem because there could be multiple ways to build the set of properties $\ias{\MI_i}$.
Furthermore, it can be the case that there are infinitely many properties in $\propertyspace_i$ that are sound, precise, and mutually incomparable.
While in this paper we do not worry about these details, the tool we use in our evaluation $\spyro$ is always guaranteed to find a maximal set of properties in $\propertyspace_i$ whenever such a set is finite (\spyro uses a regular-tree grammar to describe the set of possible properties $\propertyspace_i$, but requires such a set to be finite.)
In practice, even when the set is infinite, one can build tools that find a ``good'' set of properties and stop without
trying to find an exhaustive set.

\mypar{Discussion}
When the goal is to build a system structured in a modular fashion, modular system synthesis enables defining ``small'' synthesis problems of similar nature that concern only a single module’s implementation.

While implementation-agnostic specifications can be synthesized via the synthesis problem defined in Def.\ \ref{De:IASSynthesis}, one should be aware that there is additional flexibility to be gained if one is willing to write implementation-agnostic specifications manually.
In particular, if all of the implementation-agnostic specifications are synthesized, then it is necessary to create the system \emph{bottom-up}, synthesizing the module implementations in the order $\MI_1$, $\MI_2$, $\ldots$, $\MI_n$ (interleaved with the synthesis of $\ias{\MI_1}$, $\ias{\MI_2}$, $\ldots$, $\ias{\MI_n}$).
In contrast, when the user is willing to write the implementation-agnostic specifications manually (in addition to the implementation-specific specifications $\{ \iss{\MI_i} \}$), then the module implementations for $\MI_1$, $\MI_2$, $\ldots$, $\MI_n$ can be synthesized in any order.

%% file: 5-case-study.tex
\section{Implementation and Case-Study Evaluation}
\label{Se:ImplementationAndCaseStudy}

We carried out case studies of \framework for the simple three-layer system that has been used as a running example and for some of the modular-synthesis problems presented in the paper that introduced \jlibsketch~\cite{DBLP:journals/pacmpl/MarianoRXNQFS19}.

\subsection{Implementation}
Our implementation, called \tool, uses \jlibsketch \cite{DBLP:journals/pacmpl/MarianoRXNQFS19} to synthesize the implementation code for each layer $k$ (from the implementation-specific specification for layer $k$) and \spyro\ \cite{https://doi.org/10.48550/arxiv.2301.11117} to synthesize the implementation-agnostic specification for use at layer $k+1$.

\jlibsketch is a program-synthesis tool for Java that allows libraries to be described with collections of algebraic specifications.
Similar to its popular C counterpart \sketch~\cite{DBLP:journals/sttt/Solar-Lezama13}, \jlibsketch allows one to write programs with holes and assertions, and then tries to find integer values for the holes that cause all assertions to hold. 
Each specification is a rewrite rule of the form
$\textit{pattern} \Rightarrow \textit{result}$.
For instance, one of the rewrite rules in the specification of a stack
could be
$\pop(\push(st,k)) \Rightarrow st$.
To prevent infinite rewrite loops, a set of rewrite rules provided to \jlibsketch must not form a cycle.
For instance, the rule $a+b \Rightarrow b+a$ is not allowed.
The synthesis problem that \jlibsketch addresses is to find a program that is correct for any program input, for any library implementation that satisfies the algebraic specifications.

\lstset{%
    language=Java,
    basewidth=0.5em,
    xleftmargin=5mm,
    commentstyle=\color{dgreen}\ttfamily,
    basicstyle=\footnotesize\ttfamily,%
    numbers=left, numbersep=5pt,%
    emph={%  
    assert, harness, void, if, return, generator, int, list, boolean, minimize,%
    public, private, static, this, while, else, true, false, class,%
    rewrite, assume, stack, relation, var, new, ref%
    },emphstyle={\color{blue}}%
}%

\begin{figure}
% \vspace{-4ex}
\begin{lstlisting}
void snoc(list l, int val, ref list ret_list) {
    boolean is_empty_ret;

    ret_list = new list();
    is_empty(l, is_empty_ret);
    if (is_empty_ret) {
        ret_list.hd = val;
        nil(ret.tl);
    } else {
        ret_list.hd = l.hd;
        snoc(l.tl, val, ret.tl);
    }
}
\end{lstlisting}
\caption{Implementation of $\lsnoc$ supplied to \spyro. Returning a value from a function is done by storing the value into a reference parameter of the function.}
\label{fig:spyro-function}
\end{figure}

\spyro addresses the problem of synthesizing specifications automatically, given an implementation. 
\spyro takes as input \rone a set of function definitions $\signatures$, and \rtwo a domain-specific language $\lang$---in the form of a grammar---in which the extracted properties are to be expressed.
Properties that are expressible in $\lang$ are called \emph{\lproperties}.
\spyro outputs a set of \lproperties $\{ \phil_i \}$ that describe the behavior of $\signatures$.
Moreover, each of the $\phil_i$ is a \emph{best} \lproperty for $\signatures$:
there is no other \lproperty for $\signatures$ that is strictly more precise than $\phil_i$.
Furthermore, the set $\{ \phil_i \}$ is \emph{exhaustive}:
% the conjunction of the synthesized properties $\bigwedge_i \phil_i$ is a best \lconjunction---i.e., 
no more \lproperties can be added to it to make the conjunction $\bigwedge_i \phil_i$ more precise. 
\spyro uses \sketch as the underlying program synthesizer---i.e., it generates a number of synthesis problems in the form of \sketch files and uses \sketch to solve such problems.

Although \spyro is built on top of \sketch (instead of  \jlibsketch), in our case study we manually implemented the term-rewriting approach used by the \jlibsketch solver in the \sketch files used by \spyro to synthesize implementation-agnostic specifications that only depend on algebraic specifications of lower layers.
That is, we replace every function call $f$ appearing in a \sketch file with a function $normalize(f)$, where $normalize$ is a procedure that applies the rewrite rules from the algebraic specification.

\tool inherits the limitations of \jlibsketch and \spyro---i.e., the synthesized implementations and specifications are sound up to a bound.
Despite this limitation, the authors of \jlibsketch and \spyro have shown that these tools typically do not return unsound results in practice.
\S\ref{Se:Limitations} provides a detailed discussion of the limitations of \framework and \tool.

\subsection{Ticket-vendor Case Study}
\label{Se:CaseStudyEvaluation}

\begin{figure}
\begin{lstlisting}
var {
    int v1;
    int v2;
    list l;
    list cons_out;
    list snoc_out;
}
relation {
    cons(v1, l, cons_out);
    snoc(cons_out, v2, snoc_out);
}
generator {
    boolean AP -> !GUARD || RHS;
    boolean GUARD -> true 
                      | is_empty(l) | !is_empty(l);
    boolean RHS -> equal_list(snoc_out, L);
    int I -> v1 | v2;
    list L -> l | nil() 
              | snoc(l, I) | cons(I, L);
}
\end{lstlisting}
\caption{
Grammar for the domain-specific language
in which \spyro is to express an extracted \modlist property.
The \texttt{relation} definition in lines 8-11 specifies that the variables \texttt{snoc\_out} \texttt{l}, \texttt{v1} and \texttt{v2} are related by
$\texttt{snoc\_out} = \lsnoc(\lcons(\texttt{l}, \texttt{v1}), \texttt{v2})$.
From the grammar (``\texttt{generator}'') in lines 12-20, \spyro synthesizes best implementation-agnostic properties of form $\texttt{GUARD} \rightarrow \texttt{snoc\_out} = L$
(implicitly conjoined with $\texttt{snoc\_out} = \lsnoc(\lcons(\texttt{v1}, \texttt{l}), \texttt{v2})$).
In this case, the only expression for \texttt{GUARD} that succeeds is $\tru$, and the property synthesized is $\texttt{snoc\_out} = \lcons(\texttt{v1}, \lsnoc(\texttt{l}, \texttt{v2}))$
(with the additional implicit conjunct  $\texttt{snoc\_out} = \lsnoc(\lcons(\texttt{v1}, \texttt{l}), \texttt{v2})$).
}
\label{fig:spyro-grammar}
\end{figure}

Our first benchmark is the ticket-vending application described throughout the paper. 
% 
% In this benchmark, we have four modules \loris{todo}.
Our goal is to synthesize the four module implementations in Fig.~\ref{fig:ticket-vendor-example} (except
the bottom one), as well as the specification of each module that needs to be exposed to a higher-level module.

When synthesizing specifications, due to the scalability limitations of \spyro, we called \spyro multiple times with different smaller grammars instead of providing one big grammar of all possible properties of each module.
In each call to \spyro, we provided a grammar in which we fixed a left-hand-side expression of an equality predicate, and asked \spyro to search for a right-hand-side expression for the equality.
We allowed the right-hand-side expression to contain a conditional where the guard can be selected from the outputs of Boolean operators in the module, their negation, or constants.
For instance, 
Figures \ref{fig:spyro-function} and \ref{fig:spyro-grammar} illustrate two inputs provided to \spyro to solve the specification-synthesis problem for \modlist: \rone a program describing the implementation of \modlist (Fig.~\ref{fig:spyro-function}), and
\rtwo a grammar describing the set of possible properties (Fig.~\ref{fig:spyro-grammar}).

Because we wanted to use the synthesized equalities as input to \jlibsketch when synthesizing the implementation of the next higher-level module, we provided grammars of equalities that avoided generating cyclic rewrite rules.
We addressed this issue by limiting the search space for the right-hand-side expression.
The function symbols permitted in the right-hand-side expression are one of the functions in the left-hand-side expression,
functions used in the implementation of a function in the left-hand-side expression, or constants.
Also, the outermost function symbol of the left-hand side can only be applied to a strictly smaller term.

To illustrate some of the properties synthesized by \tool (that are not shown in Fig.~\ref{fig:ticket-vendor-example}) the complete set of equalities in the implementation-agnostic specification $\ias{\modlist}$ synthesized by \spyro is the following:
\begin{equation*}
{\small
\begin{array}{l@{\hspace{2em}}l}
    \lhead(\lcons(hd, tl)) = tl &
    \isemptylist(\lnil) = \tru \\
    \ltail(\lcons(hd, tl)) = hd &
    \isemptylist(\lcons(hd, tl)) = \fls \\
    \lsize(\lnil) = 0 &
    \lsnoc(\lnil, x) = \lcons(x, \lnil) \\
    \multicolumn{2}{l}{\lsize(\lcons(hd, tl)) = \lsize(tl) + 1} \\
    \multicolumn{2}{l}{\lsnoc(\lcons(hd, tl), x) = \lcons(hd, \lsnoc(tl, x))} \\
\end{array}
}
\end{equation*}
When considering the cumulative time taken to synthesize the algebraic specification of each module, \spyro took 41 seconds for $\ias{\modlist}$ (longest-taking property 7 seconds), 34 seconds for $\ias{\modstack}$ (longest-taking property 7 seconds), and 44 seconds for $\ias{\modqueue}$ (longest-taking property 13 seconds).

\lstset{%
    language=Java,
    basewidth=0.5em,
    xleftmargin=5mm,
    commentstyle=\color{dgreen}\ttfamily,
    basicstyle=\footnotesize\ttfamily,%
    numbers=left, numbersep=5pt,%
    emph={%  
    assert, harness, void, if, return, generator, int, list, boolean, minimize,%
    public, private, static, this, while, else, true, false, class,%
    rewrite, assume%
    },emphstyle={\color{blue}}%
}%

\begin{figure}[!t]
\begin{lstlisting}
public void enq(int x) {
    Stack st_in = this.st_in;
    Stack st_out = this.st_out;

    assume !st_out.isEmpty() || st_in.isEmpty();

    if (genGuard(st_in, st_out)) {
        st_in = genStack2(st_in, st_out, x);
        st_out = genStack2(st_in, st_out, x);
    } else {
        st_in = genStack2(st_in, st_out, x);
        st_out = genStack2(st_in, st_out, x);
    }

    assert !st_out.isEmpty() || st_in.isEmpty();

    this.st_in = st_in;
    this.st_out = st_out;
}
\end{lstlisting}
\caption{\jlibsketch sketch of $\enq$. Lines 5 and 15 
assert the implementation-specific property $\isemptystack(st_{out}) 
\rightarrow \isemptystack(st_{in})$.
\jlibsketch generates an expression to fill in
each occurrence of the generators, \texttt{genStack2} and \texttt{genGuard}---the reader can think of each of these generators as being grammars from which \jlibsketch can pick an expression.
For these generators, expressions can be variables or single function calls to functions of the appropriate type---e.g., \texttt{genStack2} can generate expressions such as \texttt{st\_in}, \texttt{st\_out}, \texttt{st\_in.pop()}, \texttt{st\_out.pop()}, etc.
}
\label{fig:enqueue-sketch}
\end{figure}

We used \jlibsketch to synthesize implementations of the modules.
In addition to the implementation-agnostic specification of the module below the one we were trying to synthesize, we provided an implementation-specific specification of the module to be synthesized.
For example, the $\iss{\modstack}$ specification involved \jlibsketch code with 17 assertions, and the following examples are an excerpt from the $\iss{\modstack}$ specification 
($x, y$, and $z$ are universally quantified integers that are allowed to be in the range 0 to 10): 
\begin{equation*}
{\small
\begin{array}{@{\hspace{0ex}}r@{\hspace{0.75ex}}c@{\hspace{0.75ex}}l@{\hspace{2.0ex}}r@{\hspace{0.75ex}}c@{\hspace{0.75ex}}l@{\hspace{0ex}}}
    \sttop(\push(\emptystack, x)) & = & x & \sttop(\push(\push(\emptystack, x), y)) & = & y \\
    \stsize(\emptystack) & = & 0 & \stsize(\push(\emptystack, x)) & = & 1 
    % \\
    % \sttop(\push(\push(\push(\emptystack, x), y), z)) = z \\
    % \sttop(\pop(\push(\push(\push(\emptystack, x), y), z))) = y \\
    % \cdots  && \cdots
\end{array}
}
\end{equation*}

Besides the assertions, we provided \jlibsketch with a fairly complete sketch of the structure of the implementation: we provided loops and branching structures, and only asked \jlibsketch to synthesize basic statements and expressions. 
For example, the sketch provided for the operation $\enq$ of module $\modqueue = (st_{in}: \modstack, st_{out}: \modstack)$ is shown in Fig.~\ref{fig:enqueue-sketch}.
This sketch of $\enq$ of module \modqueue uses two {\modstack}s: $st_{in}$, which stores elements in the rear part of the queue, and $st_{out}$, which stores elements in the front part of the queue. \modstack $st_{in}$ holds the rearmost element on top, and \modstack $st_{out}$ stores the frontmost element on top. To make the $\front$ operation more efficient, we decided to make sure that the frontmost element is always at the top of $st_{out}$. This implementation decision is expressed as assertions in lines 5 and 15, constituting an implementation-specific specification $\iss{\modqueue}$, shown as \callout{2} in Fig.~\ref{fig:ticket-vendor-example}.

Afterward, based on the implementation synthesized by \jlibsketch, \spyro was able to solve each \modqueue specification-synthesis problem within 40 seconds, yielding the following implementation-agnostic specification $\ias{\modqueue}$:
\[
{\small
\begin{array}{l}
    \isemptystack(\emptyqueue) = \tru \qquad
    \isemptyqueue(\enq(q, i)) = \fls \\
    \qsize(\emptyqueue) = 0 \\
    \qsize(\enq(q, i)) = \qsize(q) + 1 \\
    \isemptyqueue(q) \rightarrow \front(\enq(q, i)) = i \\
    \neg \isemptyqueue(q) \rightarrow \front(\enq(q, i)) = \front(q) \\
    \isemptyqueue(q) \rightarrow \deq(\enq(q, i)) = q \\
    \neg \isemptyqueue(q) \rightarrow \deq(\enq(q, i)) = \enq(\deq(q), i)\\
\end{array}
}
\]

A \modticket is implemented using a \modqueue, which stores the id numbers of clients who have reserved tickets.
Each issued ticket contains the id of the buyer.
The implementation-specific specification $\iss{\modticket}$ consisted of \jlibsketch code with 24 assertions, and contains multiple examples, such as the following (again, $x$ and $y$ are universally quantified integers that are allowed to be in the range 0 to 10): 
\[
{\small
\begin{array}{l}
    \begin{array}{l}
    \numticket(\preparesales(2)) = 2 \\
    \numwaiting(\preparesales(2)) = 0 \\
    \numwaiting(\reserveticket(\preparesales(2), x)) = 1 \\
    \issueticket(\reserveticket(\preparesales(2), x)).owner = x \\
    % \issueticket(\reserveticket(\reserveticket(\preparesales(2), x), y)).owner = y \\
    % \ldots
    \end{array}
\end{array}
}
\]

Again, we provided \jlibsketch with a fairly complete sketch of the program structure, and \jlibsketch was able to synthesize the implementations of all the \modticket functions within 10 seconds. 
For example, the function $\preparesales$ for $\modticket = (num_{ticket}: \inttype, q_{waiting}: \modqueue)$ was synthesized as $\preparesales(n: \inttype) := (n, \emptyqueue)$.

We compared the time needed to synthesize each module from the algebraic specification of the previous module to the time needed to synthesize using the implementation of all previous modules.
Synthesizing \modstack from the specification $\ias{\modlist}$ took 3 seconds instead of the 2 seconds needed when the implementation of \modlist was provided.
Synthesizing \modqueue from the specification $\ias{\modstack}$ took 188 seconds instead of the 799 seconds needed when the concrete implementations of \modstack and \modlist were provided.
Synthesizing \modticket from the specification $\ias{\modqueue}$ took 7 seconds, but \jlibsketch crashed when the concrete implementations of \modqueue, \modstack and \modlist were provided.

\textbf{Key finding:}
This experiment shows that modular synthesis takes 1-5 minutes per module, whereas the time taken to synthesize a module from the underlying module implementations grows with the number of modules---to the point where synthesis is unsuccessful with existing tools.

As discussed in \S\ref{Se:DiffQueue}, we also synthesized an implementation of \modqueue that uses \modlist instead of two {\modstack}s.
The \modlist holds the oldest element of the \modqueue at its head. 
The implementation-specific specification $\iss{\modqueue~(as~\modlist)}$ consisted of \jlibsketch code with 19 assertions, including examples similar to those shown at
\callout{5} in Fig.~\ref{fig:ticket-vendor-example/queuelist}.
We used \jlibsketch to verify whether the specification $\ias{\modqueue}$ still held true for the new implementation.
Because it did (confirmation took ${<} 1$ second),
\modticket does not need to be changed to use the \modqueue~(as~\modlist) implementation.

\subsection{Case Studies from Mariano et al.~\cite{DBLP:journals/pacmpl/MarianoRXNQFS19}}
\label{Se:CaseStudyEvaluationMariano}

Our second set of benchmarks is collected from the paper that introduced synthesis from algebraic specifications via \jlibsketch~\cite{DBLP:journals/pacmpl/MarianoRXNQFS19}.
In that work, Mariano et al.\ used a number of benchmarks that involve two modules---e.g., synthesizing a backend cryptographic component for a tool
that brings NuCypher to Apache Kafka, using \modarraylist and \modhm as underlying modules.
The goal of their paper was to show that in \jlibsketch it was easier/faster to synthesize the module at layer 1 when the module of layer 0 was exposed through an algebraic specification (rather than a concrete implementation).
The current implementation of \tool does not support strings, so we used only the benchmarks for which the algebraic specifications for the layer-0 module (i) did not use \texttt{string} operations, and (ii) did not use auxiliary functions that were not in the signature of the module.
In total, we considered four layer-0 modules: \modarraylist, \modts, \modhs, and \modhm.
Each \jlibsketch benchmark consisted of (i) an algebraic specification of the layer-0 module (written by hand), (ii) a \sketch-like specification of the layer-1 module, and (iii) a mock implementation of the layer-0 module---i.e., a simplified implementation that mimics the module's intended behavior
(e.g., \texttt{HashSet} is implemented using an array).
  The mock is not needed by \jlibsketch, but allowed Mariano et al. to compare synthesis-from-algebraic-specifications against synthesis-from-mocks \cite[\S5]{DBLP:journals/pacmpl/MarianoRXNQFS19}.
 
We used these items in a different manner from the \jlibsketch experiments.
From just the mock implementation of layer 0, we asked \tool to synthesize a most-precise algebraic specification, which we compared with the algebraic specification manually written by Mariano et al.
From that algebraic specification and the \sketch-like specification of the layer-1 module, we asked \tool to synthesize the implementation of layer 1.
(The second step essentially replicated the algebraic-synthesis part of the \jlibsketch experiments.) 

For the layer-0 synthesis step of each benchmark, we synthesized algebraic specifications using grammars similar to the ones used in \S\ref{Se:CaseStudyEvaluation}.

When considering the time taken to synthesize the entire algebraic specification of each module, \spyro took 626 seconds for $\ias{\modarraylist}$, 54 seconds for $\ias{\modhs}$, and 1,732 seconds for $\ias{\modhm}$.
Because mock implementations are simplified versions of actual implementations, the mock implementation of $\modts$ is identical to the mock implementation of $\modhs$---i.e., they both represent sets as arrays.
Furthermore, the two implementations have the same algebraic specifications---i.e., $\ias{\modhs}=\ias{\modts}$---which can thus be synthesized in the same amount of time.

\textbf{Key finding:} For all but two benchmarks, the \lconjunctions synthesized by \tool were equivalent to the algebraic properties manually written by Mariano et al.
For the mock implementation of \texttt{\modhm} and \modarraylist provided in \jlibsketch, for specific grammars, \tool synthesized empty \lconjunctions (i.e., the predicate \textit{true}) instead of the algebraic specifications provided by Mariano et al.---i.e.,
$\kone = \ktwo \Rightarrow \mapget(\mapput(\varm, \kone, \varval), \ktwo) = \varval$ and 
$\vari = \varj \Rightarrow \arrayget(\arrayset(\varl, \vari, \varval), \varj) = \varval$, for \texttt{\modhm} and \texttt{\modarraylist}, respectively.
Upon further inspection, we discovered that \jlibsketch's mock implementation of \texttt{\modhm} was incorrect, and did not satisfy the specification that Mariano et al.\  gave, due to an incorrect handling of hash collision!
After fixing the bug in the mock implementation of \modhm, we were able to synthesize the expected algebraic specification.
However, when inspecting the implementation of \modarraylist, we found that for this benchmark the implementation was correct but the algebraic specification provided by Mariano et al.\ was incorrect!
After modifying the grammar, we could synthesize the correct algebraic specification
$(\vari = \varj) \wedge (0 \leq \vari) \wedge (\vari \leq \lsize(\varl)) \Rightarrow \arrayget(\arrayset(\varl, \vari, \varval), \varj) = \varval$.
However, this modification revealed a bug in one of the implementations of \modhm that Mariano et al.\ had synthesized from the earlier erroneous specification!
We discuss this finding further in the next section.

This finding illustrates how modular system synthesis can help to \textit{identify} and \textit{avoid} bugs in module implementations.

\subsection{Additional Case Studies Based on Mariano et al.~\cite{DBLP:journals/pacmpl/MarianoRXNQFS19}}
\label{Se:AdditionalCaseStudyEvaluationMariano}
We noticed that the \jlibsketch benchmarks provided an opportunity to build a more complicated benchmark that involved 3 modules (instead of 2).
In particular, two of the benchmarks involved synthesizing the implementation of a (layer-1) \modhm module from a (layer-0) algebraic specification of \modarraylist. 
(The two benchmarks synthesized different implementations that handled collisions differently and we refer to the corresponding modules as \modhmone and \modhmtwo.)
The third benchmark involved synthesizing the implementation of a (layer-2) \modkafka from a (layer-1) algebraic specification of \modhm.
Thus, we built two 3-layer benchmarks in which the goal was to synthesize \modkafka using an implementation of \modhm that used an implementation of \modarraylist.
For us, each 3-layer benchmark involved four synthesis problems:
(1) the algebraic specification $\ias{\modarraylist}$ of \modarraylist (from the mock);
(2) the implementation of either \modhmone or \modhmtwo;
(3) the algebraic specification of \modhm; and
(4) the implementation of \modkafka (this part was already synthesized in~\cite{DBLP:journals/pacmpl/MarianoRXNQFS19}).

As discussed in the previous section, we identified a bug in the specification $\ias{\modarraylist}$ manually provided by Mariano et al., and were able to use to \tool to synthesize a correct algebraic specification---i.e., step (1).
For step (2), the implementation synthesized by Mariano et al.\ for \modhmtwo was still correct, and we could also use \tool to synthesize it from the corrected specification $\ias{\modarraylist}$.
However, the implementation of \modhmone synthesized by \jlibsketch was incorrect because it depended on the original, erroneous specification $\ias{\modarraylist}$ for \modarraylist---(1) $\mapput$ could store values to negative indices; and (2) $\mapget$ could search key from incorrect index after rehashing.
We manually changed the implementation of the rehashing function in the sketch of \modhmone to fix the bug, but the change was large enough that we did not attempt to rewrite the program sketch needed to synthesize this specification (i.e., we manually wrote the implementation of \modhmone instead of synthesizing it). 
Synthesis problem (3) is at the heart of handling a multi-module system in a modular fashion: we used \tool to synthesize algebraic specifications of \modhmone and \modhmtwo---in each case, giving \tool access to the (correct) implementations of \modhmone and \modhmtwo and the (correct) algebraic specification of \modarraylist (but not an implementation of \modarraylist). 

\textbf{Key finding:}  
\tool failed to synthesize the same algebraic specification we had obtained for \modhm in \S\ref{Se:CaseStudyEvaluationMariano} when attempting to synthesize a specification for \modhmone and \modhmtwo.
When inspecting the synthesized properties, we realized that the algebraic specification $\ias{\modarraylist}$ exposed by \modarraylist still had a problem!
In particular, $\ias{\modarraylist}$ was too weak to prove the algebraic specifications needed by \modhmone and \modhmtwo---i.e., $\ias{\modarraylist}$ did not characterize properties that were needed by \modhmone and \modhmtwo to satisfy the algebraic specification $\ias{\modhm}$.
We used Sketch itself to produce a violation of the algebraic specification $\ias{\modhm}$ for \modhmone under the weaker assumption that \modarraylist only satisfied the specification $\ias{\modarraylist}$, and used the violations generated by \sketch to identify what properties we needed to add to strengthen $\ias{\modarraylist}$.
In particular, $\lsize(\arrayec(\varl, \varn)) = \lsize(\varl)$ and $\arrayget(\arrayec(\varl, \varn), \vari) = \arrayget(\varl, \vari)$ were added to describe the behavior of $\arrayec$.
We were then able to modify the grammar used to synthesize algebraic specifications for $\ias{\modarraylist}$ and synthesize the missing property.
After obtaining $\ias{\modarraylist}$, we successfully synthesized the full algebraic specification for $\modhmtwo$ (i.e., $\ias{\modhm}$) and most of the algebraic specification for $\modhmone$.
Because the corrected implementation of $\modhmone$ was particularly complicated---e.g.,
each call to $\mapput$ requires rehashing when the load factor is greater than a predefined value---\tool timed out while synthesizing every property, with the exception of the property $\mapget(\mapempty, \vark) = \error$.

This finding illustrates how modular system synthesis can help identify  when module specifications are not strong enough to characterize the behavior of other modules.

\subsection{Limitations of \tool}
\label{Se:Limitations}

\jlibsketch and \spyro represent the algebraic specifications of modules as rewrite rules for algebraic datatypes (ADTs). Reasoning about ADTs is a challenging problem, and to the best of our knowledge, \sketch and \jlibsketch are only frameworks capable of handling problems involving ADTs effectively. Therefore, \tool uses them as the underlying solver and inherits limitations of \sketch.

The primary limitation of \tool is its bounded soundness guarantee.
\sketch ensures soundness only for a bounded number of loop/recursion unrollings, and bounded input sizes. Verifying the unbounded correctness of the synthesized programs poses a significant challenge, as semantics of lower-level modules are represented as rewrite rules on ADTs. As a future direction, we plan to integrate \tool with verifiers such as Dafny to perform full verification, as was done in \cite{https://doi.org/10.48550/arxiv.2301.11117} for the properties synthesized by \spyro.
However, it is worth noting that \tool has already been useful in finding bugs in existing implementations: specification synthesis has helped find implementation errors in the case studies of Mariano et al. \cite{DBLP:journals/pacmpl/MarianoRXNQFS19}, as demonstrated in \S\ref{Se:CaseStudyEvaluationMariano} and \S\ref{Se:AdditionalCaseStudyEvaluationMariano}.

Although the case studies in \S\ref{Se:CaseStudyEvaluation} and reference \cite{DBLP:journals/pacmpl/MarianoRXNQFS19} show satisfactory performance of \sketch for most problems, scalability issues persist.
In particular, unrolling nested loops significantly increases the number of holes of the \sketch problem, which increases the problem's difficulty.

% \twr{Can't the modular structure of a system be a dag?
% Also, not sure why we are saying ``bottom-up'' here?
% I think two points are being mixed: (i) the modular structure is tree-shaped or dag-shaped, versus (ii) one convenient order in which to synthesize modules is to work bottom-up.}
Besides the limitations inherited from \sketch, \framework has a specific requirement for the system's modular structure, which should be a directed acyclic graph (DAG)---i.e., the implementation-agnostic specifications of all dependent modules must be provided to synthesize a particular module.
\framework addresses the challenges in writing accurate specifications by using the synthesis of implementation-agnostic specifications. 
However, in this approach one needs to synthesize all dependent modules and their specifications before attempting to synthesize a new module. 
Alternatively, to synthesize higher-level modules without the lower-level implementations, the user can manually supply the implementation-agnostic specifications of the lower-level modules.

%% file: 6-related-work.tex
\section{Related Work}
\label{Se:RelatedWork}

A problem related to ours is that of component-based synthesis (CBS), where the goal is \textit{assembling} pre-existing components/APIs to generate more complex programs.
    Many existing approaches for solving CBS problems scale reasonably well~\cite{DBLP:conf/popl/FengM0DR17,DBLP:journals/pacmpl/ShiSL19,singh14}, but require the individual components to be executable.
    In our setting, this approach is not possible because
    the details of lower-level components (e.g., how a \modstack is implemented) need not be observable. 

A few tools have abstracted components and modules using specifications. 
\jlibsketch~\cite{DBLP:journals/pacmpl/MarianoRXNQFS19} uses algebraic properties to represent the semantics of modules and is a key component of our implementation.
(CL)S~\cite{isola} and APIphany~\cite{guo22} use types to represent the behavior of components and can be used in tandem with specialized type-directed synthesizers.
The key differences between our work and these tools is that \framework provides two well-defined synthesis primitives that support composing multiple modules, rather than synthesizing just one implementation for one module.
Furthermore, the aforementioned types are limited in how they can represent relations between multiple components in an implementation-agnostic way, thus making us opt for algebraic specifications.

    Many synthesis tools perform some kind of ``compositional'' synthesis by breaking an input specification into sub-specifications that are used to separately synthesize sub-components of a target program~\cite{ijcai15gulwani,DBLP:conf/cav/AlurCR15}.
    This notion of ``compositionality'' is orthogonal to ours,
    and is more of a divide-and-conquer approach to solving \emph{individual} synthesis problems.
    \framework can make use of such a divide-and-conquer approach when synthesizing a module's implementation.

    For the task of synthesizing an algebraic specification, \tool uses \spyro. 
    Besides \spyro, there are a number of works about discovering specifications from code, based on both static techniques~\cite{DBLP:conf/fm/FlanaganL01,DBLP:journals/corr/abs-1905-06847} and dynamic techniques~\cite{DBLP:journals/scp/ErnstPGMPTX07,DBLP:journals/tse/HenkelRD07}.
    The static approaches mostly target predicates involving individual functions (instead of algebraic properties and equalities involving multiple functions).
    The dynamic techniques are flexible and can identify algebraic specifications (e.g., for Java container classes~\cite{DBLP:journals/tse/HenkelRD07}), but require some ``bootstrapping'' inputs, and only guarantee soundness with respect to behaviors that are covered by the tests that the inputs exercise.

%% file: 7-conclusion.tex
\section{Conclusion}
\label{Se:Conclusion}

% This paper (i) defines modular system synthesis;
% (ii) illustrates the concepts by means of several examples; and
% (iii) describes a proof-of-concept system, called \tool, that achieves these goals.

\mypar{Conceptual contributions}
At the conceptual level, this paper contributes both a framework and a new way to think about program synthesis that opens many research directions.
Specifically, the paper introduces \framework, a framework for using synthesis to perform modular system synthesis.
The main contribution of this paper is not an immediate solution to the modular-synthesis problem, but rather the identification of two key synthesis primitives that are required to realize \framework in practice:
1) synthesis \underline{from} an implementation-agnostic specification, and
2) synthesis \underline{of} an implementation-agnostic specification.
While our tool implements both of these primitives using tools based on \sketch (thus inheriting its limitations), an interesting research directions is whether other synthesis approaches (enumeration, CEGIS, etc.) can be extended to handle our synthesis problems, perhaps by leveraging the popular \texttt{egg} framework~\cite{DBLP:journals/pacmpl/WillseyNWFTP21}\, which allows one to reason about equivalence of terms with respect to a term-rewriting system---i.e., our algebraic specifications.

\mypar{Experimental Contributions}
We created \tool, a proof-of-concept implementation of \framework based on two existing program-synthesis
tools:
\jlibsketch~\cite{DBLP:journals/pacmpl/MarianoRXNQFS19}, a program-sketching tool that supports algebraic specifications, and \spyro~\cite{https://doi.org/10.48550/arxiv.2301.11117}, a tool for synthesizing precise specifications from code.
The case studies carried out with \tool show that \rone modular synthesis is faster than monolithic synthesis, and \rtwo performing synthesis for both implementations and specifications of the modules can prevent subtle bugs.    

\section*{Acknowledgement}

Supported, in part,
by a Microsoft Faculty Fellowship, a gift from Rajiv and Ritu Batra; by ONR under grant N00014-17-1-2889; and by NSF under grants CCF-\{1750965,1763871,1918211,2023222,2211968,2212558\}.
Any opinions, findings, and conclusions or recommendations expressed in this publication are those of the authors, and do not necessarily reflect the views of the sponsoring entities.

%% file: 8-appendix-detailed-case-study.tex
\subsection{Ticket-vendor Detailed Case Study}
\label{app:ticket-vendor}

In \tool, to synthesize the implementation-agnostic specification of the operations $\MI_k$ in layer $k$, we supplied \spyro with the code corresponding to the implementations of the functions $\MI_k$, and a domain-specific language $\lang$ of equalities over the functions $\MI_k$. 
Although \spyro is built on top of \sketch (instead of  \jlibsketch), 
we manually implemented the term rewriting approach of \jlibsketch in the \sketch files used by \spyro in our case study to synthesize implementation-agnostic specifications that only depend on algebraic specifications of lower layers. 

\mypar{{\normalfont\modlist} Specification Synthesis}
As shown in Fig.~\ref{fig:ticket-vendor-example}, we assumed that \spyro, used with a specific implementation of \modlist, synthesized an implementation-agnostic specification for operations in $\PL[\modlist]$---i.e., $\lnil$, $\lcons$, $\lhead$, $\ltail$, $\lsnoc$, $\lsize$, and $\isemptylist$.
Due to the current scalability limitations of \spyro, we called \spyro multiple times with different smaller grammars instead of providing one big grammar of all possible properties.
In each call to \spyro, we provided a grammar in which we fixed a left-hand-side expression of an equality predicate, and asked \spyro to search for a right-hand-side expression for the equality.
We allowed the right-hand-side expression to contain a conditional where the guard can be selected from the outputs of Boolean operators in the module, their negation, or constants.

Because we wanted to use the synthesized equalities as input to \jlibsketch when synthesizing implementations for the \modstack module, we provided grammars of equalities that avoided generating cyclic rewrite rules.
We addressed this issue by limiting the search space for the right-hand-side expression.
The function symbols permitted in the right-hand-side expression are one of the functions in the left-hand-side expression,
functions used in the implementation of a function in the left-hand-side expression, or constants.
Also, the outermost function symbol of the left-hand side can only be applied to a strictly smaller term.
For instance, in one of the calls to \spyro, the goal is to find values of $\textit{guard}$ and $\textit{exp}$ that satisfy the following equation:
\begin{equation}
    \textit{guard} \rightarrow \lsnoc(\lcons(hd, tl), x) = \textit{exp}
\label{eq:apx-list-spec-problem}
\end{equation}
where $\textit{guard}$ is one of $\isemptylist(l)$, $\neg \isemptylist(l)$ or $\tru$, and $\textit{exp}$ is expressed by the grammar
$L  :=  tl \mid \lnil \mid \lsnoc(tl, I) \mid \lcons(I, L); I :=  hd \mid x$.

\spyro was able to solve each \modlist specification-synthesis problem within 10 seconds. 
For the problem in Eq.~(\ref{eq:apx-list-spec-problem}), \spyro synthesized $\textit{guard} = \tru$ and $\textit{exp} = \lcons(hd, \lsnoc(tl, x))$.
The complete set of equalities in the implementation-agnostic specification $\ias{\modlist}$ synthesized by \spyro is the following:
\[
\begin{array}{lcl}
    \isemptylist(\lnil) = \tru \qquad
    \isemptylist(\lcons(hd, tl)) = \fls \\
    \lsize(\lnil) = 0 \\
    \lsize(\lcons(hd, tl)) = \lsize(tl) + 1 \\
    \lhead(\lcons(hd, tl)) = tl \\
    \ltail(\lcons(hd, tl)) = hd \\
    \lsnoc(\lnil, x) = \lcons(x, \lnil) \\
    \lsnoc(\lcons(hd, tl), x) = \lcons(hd, \lsnoc(tl, x)) \\
\end{array}
\]

\mypar{{\normalfont\modstack} Implementation Synthesis}
We then used \jlibsketch to synthesize an implementation of the \modstack operations $\emptystack$, $\push$, $\sttop$, $\pop$, $\stsize$, and $\isemptystack$.
In this implementation, a \modstack uses a \modlist.
When building the \jlibsketch files for this step, we manually translated the implementation-agnostic specification $\ias{\modlist}$ synthesized by \spyro in the previous step into \jlibsketch rewrite rules. 

On top of the implementation-agnostic specification of the \modlist module, we also provided an implementation-specific specification $\iss{\modstack}$ for the kind of \modstack we were trying to synthesize. 
The $\iss{\modstack}$ specification involved \jlibsketch code with 17 assertions.
The following examples are an excerpt from the $\iss{\modstack}$ specification 
($x, y$, and $z$ are universally quantified integers that are allowed to be in the range 0 to 10): 
\[
\begin{array}{lcl}
    \sttop(\push(\emptystack, x)) = x \\
    \stsize(\emptystack) = 0 \\
    \sttop(\push(\push(\emptystack, x), y)) = y \\
    \stsize(\push(\emptystack, x)) = 1 
    % \\
    % \sttop(\push(\push(\push(\emptystack, x), y), z)) = z \\
    % \sttop(\pop(\push(\push(\push(\emptystack, x), y), z))) = y \\
    % \cdots  && \cdots
\end{array}
\]

Besides the assertions, we provided \jlibsketch with a fairly complete sketch of the structure of the implementation: we provided
loops and branching structures and only asked \jlibsketch to synthesize basic statements and expressions. \jlibsketch was able to synthesize the implementations of all the \modstack functions within 10 seconds. 
For example, the function $\pop$ for $\modstack = (l: \modlist)$ was synthesized as $\pop(st: \modstack) := \ltail(st.l)$.

\mypar{{\normalfont\modstack} Specification Synthesis}
Our implementation-specific specification $\iss{\modstack}$ does not contain any function symbols from $\PL[\modlist]$---i.e., it was actually implementation-agnostic.
However, since $\iss{\modstack}$ only describes the behavior for specific examples, we used \spyro to synthesize a new implementation-agnostic specification of \modstack that generalized to arbitrary inputs.
To use \spyro, we manually translated the \modstack implementation computed by \jlibsketch into code that could be used by \spyro.

By providing grammars similar to the ones provided
for the \modlist functions for the \modlist specification-synthesis problem,
\spyro was able to solve each \modstack specification-synthesis problem within 30 seconds, and computed 
 the implementation-agnostic specification $\ias{\modstack}$ presented in Fig.~\ref{fig:ticket-vendor-example} in \S\ref{Se:IllustrativeExample}.
% \[
% \begin{array}{c@{\hspace{2ex}}c}
% \isemptystack(\emptystack) = \tru & \isemptystack(\push(st, x)) = \fls \\
% \stsize(\emptystack) = 0 & \stsize(\push(st, x)) = \stsize(st) + 1 \\
% \pop(\push(st, x)) = st & \sttop(\push(st, x)) = x
% \end{array}
% \]

\mypar{{\normalfont\modqueue} Implementation Synthesis}
We then used \jlibsketch to synthesize an implementation of the \modqueue operations $\emptyqueue$, $\enq$, $\front$, $\deq$, $\qsize$, and $\isemptyqueue$.
A \modqueue is implemented using two {\modstack}s: $st_{in}$, which stores elements in the rear part of the queue, and $st_{out}$, which stores elements in the front part of the queue. \modstack $st_{in}$ holds the rearmost element on top, and \modstack $st_{out}$ stores the frontmost element on top. To make the $\front$ operation more efficient, we decided to make sure that the frontmost element is always at the top of $st_{out}$. 
% The implementation-agnostic specifications of \modstack are manually translated into \jlibsketch rewrite rules.

The implementation-specific specification $\iss{\modqueue}$ for the \modqueue operations consisted of \jlibsketch code with 20 assertions.
The assertions included invariants relating the two stacks, such as 
$\isemptystack(st_{out}) \rightarrow \isemptystack(st_{in})$,
as well as such examples as 
\[
\begin{array}{lcl}
    \front(\enq(\emptyqueue, x)) = x \\
    \qsize(\emptyqueue) = 0 \\
    \front(\enq(\enq(\emptyqueue, x), y)) = y \\
    \qsize(\enq(\emptyqueue, x)) = 1 
    % \front(\enq(\enq(\enq(\emptyqueue, x), y), z)) = z \\
    % \front(\deq(\enq(\enq(\enq(\emptyqueue, x), y), z))) = y \\
    % \cdots \\
    % \cdots
\end{array}
\]
Again, $x, y$, and $z$ are universally quantified integers that are allowed to be in the range 0 to 10.
Again, we provided \jlibsketch with a fairly complete sketch of the program structure, and \jlibsketch was able to synthesize all the \modqueue implementations within 10 seconds. 
For example, the function $\enq$ for $\modqueue = (st_{in}: \modstack, st_{out}: \modstack)$ was synthesized as $\enq(q: \modqueue) := 
\mathtt{if}\; \isemptystack(st_{out})\;$
$\mathtt{then}\; (q.st_{in}, \push(q.st_{out}, i))\;$
$\mathtt{else}\; (\push(q.st_{in}, i), q.st_{out})$.
This implementation is correct due to the invariant $\isemptystack(st_{out}) \rightarrow \isemptystack(st_{in})$, because this property ensures that $st_{out}$ is empty only if both stacks $st_{in}$ and $st_{out}$ are empty.

\mypar{{\normalfont\modqueue} Specification Synthesis}
With an experimental setup similar to the one for \modstack specification synthesis, \spyro was able to solve each \modqueue specification-synthesis problem within 40 seconds, yielding the following implementation-agnostic specification $\ias{\modqueue}$:
\[
\begin{array}{l}
    \isemptystack(\emptyqueue) = \tru \qquad
    \isemptyqueue(\enq(q, i)) = \fls \\
    \qsize(\emptyqueue) = 0 \\
    \qsize(\enq(q, i)) = \qsize(q) + 1 \\
    \isemptyqueue(q) \rightarrow \front(\enq(q, i)) = i \\
    \neg \isemptyqueue(q) \rightarrow \front(\enq(q, i)) = \front(q) \\
    \isemptyqueue(q) \rightarrow \deq(\enq(q, i)) = q \\
    \neg \isemptyqueue(q) \rightarrow \deq(\enq(q, i)) = \enq(\deq(q), i)\\
\end{array}
\]
\mypar{{\normalfont\modticket} Implementation Synthesis}
We used \jlibsketch to synthesize an implementation of the \modticket operations $\reserveticket$, $\issueticket$, $\soldout$, $\numticket$, and $\numwaiting$.
A \modticket is implemented using a \modqueue, which stores the id numbers of clients who have reserved tickets.
Each issued ticket contains the id of the buyer.

The implementation-specific specification $\iss{\modticket}$ consisted of \jlibsketch code with 24 assertions, and contains multiple examples, such as the following (again, $x$ and $y$ are universally quantified integers that are allowed to be in the range 0 to 10): 
\[
\begin{array}{l}
    \begin{array}{l}
    \numticket(\preparesales(2)) = 2 \\
    \numwaiting(\preparesales(2)) = 0 \\
    \numwaiting(\reserveticket(\preparesales(2), x)) = 1 \\
    \issueticket(\reserveticket(\preparesales(2), x)).owner = x \\
    % \issueticket(\reserveticket(\reserveticket(\preparesales(2), x), y)).owner = y \\
    % \ldots
    \end{array}
\end{array}
\]

Again, we provided \jlibsketch with a fairly complete sketch of the program structure, and \jlibsketch was able to synthesize the implementations of all the \modticket functions within 10 seconds. 
For example, the function $\preparesales$ for $\modticket = (num_{ticket}: \inttype, q_{waiting}: \modqueue)$ was synthesized as $\preparesales(n: \inttype) := (n, \emptyqueue)$.

\mypar{Changing the \modqueue Implemenation}
As illustrated in \S\ref{Se:DiffQueue}, we also synthesized a different implementation of \modqueue that uses \modlist instead of two {\modstack}s. The \modlist holds the oldest element of the \modqueue at its head. 
The implementation-specific specification $\iss{\modqueue~(as~\modlist)}$ consisted of \jlibsketch code with 19 assertions, including such examples as 
\[
\begin{array}{lcl}
    \front(\enq(\emptyqueue, x)) = x \\
    \qsize(\emptyqueue) = 0 \\
    \front(\enq(\enq(\emptyqueue, x), y)) = y \\
    \qsize(\enq(\emptyqueue, x)) = 1 
    % \front(\enq(\enq(\enq(\emptyqueue, x), y), z)) = z \\
    % \front(\deq(\enq(\enq(\enq(\emptyqueue, x), y), z))) = y \\
    % \cdots \\
    % \cdots
\end{array}
\]
where $x, y$ and $z$ are any distinct integers between 0 and 10.

Because we synthesized the implementation-agnostic specification $\ias{\modqueue}$ from 
the previous implementation, as a sanity check we used \jlibsketch to verify whether the specification $\ias{\modqueue}$ still held true for the new implementation. Because this was the case (the check took less than a second), \modticket does not need to be changed to use the \modqueue-as-\modlist implementation.

%% file: 9-appendix-code.tex
\subsection{Implementation Synthesis with JLibSketch}
\label{app:implementation-synthesis}

We present the three inputs provided to \jlibsketch to solve the implementation-synthesis problem for \modqueue: \rone a program sketch describing the search space of possible programs (Fig.~\ref{fig:jlibsketch-sketch}),
\rtwo an implementation-agnostic specification $\ias{\modstack}$ of the \modstack module in the form of rewrite rules (Fig.~\ref{fig:jlibsketch-rewrite-class}), and
\rthree an implementation-specific specification $\iss{\modqueue}$ of the \modqueue module in the form of assertions (Fig.~\ref{fig:jlibsketch-test}).

\lstset{%
    language=Java,
    basewidth=0.5em,
    xleftmargin=5mm,
    commentstyle=\color{dgreen}\ttfamily,
    basicstyle=\footnotesize\ttfamily,%
    numbers=left, numbersep=5pt,%
    emph={%  
    assert, harness, void, if, return, generator, int, list, boolean, minimize,%
    public, private, static, this, while, else, true, false, class,%
    rewrite, assume%
    },emphstyle={\color{blue}}%
}%

\begin{figure}[!t]
\begin{lstlisting}
public void enqueue(int x) {
    Stack st_in = this.st_in;
    Stack st_out = this.st_out;

    assume !st_out.isEmpty() || st_in.isEmpty();

    if (genGuard(st_in, st_out)) {
        st_in = genStack2(st_in, st_out, x);
        st_out = genStack2(st_in, st_out, x);
    } else {
        st_in = genStack2(st_in, st_out, x);
        st_out = genStack2(st_in, st_out, x);
    }

    assert !st_out.isEmpty() || st_in.isEmpty();

    this.st_in = st_in;
    this.st_out = st_out;
}

private static void rev(Stack in, Stack out) {
    while(!in.isEmpty()) {
        out.push(in.top());
        in.pop();            
    }
}

public void dequeue() {
    Stack st_in = this.st_in;
    Stack st_out = this.st_out;

    assume !st_out.isEmpty() || st_in.isEmpty();

    st_in = genStack1(st_in, st_out);
    st_out = genStack1(st_in, st_out);

    if (genGuard(st_in, st_out)) {
        rev(st_in, st_out);
    }

    this.st_in = st_in;
    this.st_out = st_out;

    assert !st_out.isEmpty() || st_in.isEmpty();
}
\end{lstlisting}
\caption{\jlibsketch sketch of $\enq$ and $\deq$. Line 5, 15, 32 and 44 
assert the implementation-specific property $\isemptystack(st_{out}) 
\rightarrow \isemptystack(st_{in})$.
\jlibsketch generates an expression to fill in
each occurrence of the generators \texttt{genStack1}, \texttt{genStack2} and \texttt{genGuard}---the reader can think of each of these generators as being grammars from which \jlibsketch can pick an expression.
For these generators, expressions can be variables or single function calls to functions of the appropriate type---e.g., \texttt{genStack1} can generate expressions such as \texttt{st\_in}, \texttt{st\_out}, \texttt{st\_in.pop()}, \texttt{st\_out.pop()}, etc.
}
\label{fig:jlibsketch-sketch}
\end{figure}

\newpage

\begin{figure}
\begin{lstlisting}
@rewriteClass
class Stack {
    @alg
    Stack push(int x);

    @alg
    @pure
    int top();

    @alg
    Stack pop();

    @alg
    @pure
    int size();

    @alg
    @pure
    boolean isEmpty();

    rewrite int size(Stack Stack()) { return 0; }
    rewrite int size(
                Stack push!(Stack st, int x)) { 
        return size(st) + 1; 
    }
    rewrite boolean isEmpty(Stack Stack()) { 
        return true; }
    rewrite boolean isEmpty(
                    Stack push!(Stack st, int x)) { 
        return false; 
    }
    rewrite int top(
                Stack push!(Stack st, int x)) { 
        return x; }
    rewrite Stack pop!(
                Stack push!(Stack st, int x)) { 
        return st; }
}
\end{lstlisting}
\caption{\jlibsketch rewrite class \modstack for the synthesis of \modqueue. Lines 3-19 are function signatures of \modstack operations, and lines 22-31 are implementation-agnostic properties $\ias{\modstack}$ of \modstack.
The constructor \modstack{()} plays the same role as what was referred to in the body of the paper as $\emptystack$.
}
\label{fig:jlibsketch-rewrite-class}
\end{figure}

\newpage

\begin{figure}
\begin{lstlisting}
harness void test(int x, int y, int z) {
    assume x != y && x != z && y != z;
    assume x > 0 && x < 10;
    assume y > 0 && y < 10;
    assume z > 0 && z < 10;

    Queue queueUnderTest = Queue.empty();
    // size_q(empty_q) == 0
    assert queueUnderTest.size() == 0;
    // is_empty_q(empty_q) == true
    assert queueUnderTest.isEmpty();

    queueUnderTest.enqueue(x);
    // size_q(enqueue(empty_q,x)) == 1
    assert queueUnderTest.size() == 1;
    // front(enqueue(empty_q,x)) == x
    assert queueUnderTest.front() == x;
    // is_empty_q(enqueue(empty_q,x)) == false
    assert !queueUnderTest.isEmpty();

    queueUnderTest.enqueue(y);
    assert queueUnderTest.size() == 2;
    assert queueUnderTest.front() == x;
    assert !queueUnderTest.isEmpty();

    queueUnderTest.enqueue(z);
    assert queueUnderTest.size() == 3;
    assert queueUnderTest.front() == x;
    assert !queueUnderTest.isEmpty();

    queueUnderTest.dequeue();
    assert queueUnderTest.size() == 2;
    assert queueUnderTest.front() == y;
    assert !queueUnderTest.isEmpty();

    queueUnderTest.dequeue();
    assert queueUnderTest.size() == 1;
    assert queueUnderTest.front() == z;
    assert !queueUnderTest.isEmpty();
}
\end{lstlisting}
\caption{\jlibsketch harness corresponding to the implementation-specific specification $\iss{\modqueue}$ for \modqueue operations. Lines 2-5
specify
a range of integers to be tested, and lines 7-39 checks the behavior of various functions using a specific test cases.
For a few property, we include comments describing what property is being tested.
}
\label{fig:jlibsketch-test}
\end{figure}

%% file: main.bbl
\begin{thebibliography}{10}

\bibitem{DBLP:conf/cav/AlurCR15}
R.~Alur, P.~Cern{\'{y}}, and A.~Radhakrishna.
\newblock Synthesis through unification.
\newblock In D.~Kroening and C.~S. Pasareanu, editors, {\em Computer Aided
  Verification - 27th International Conference, {CAV} 2015, San Francisco, CA,
  USA, July 18-24, 2015, Proceedings, Part {II}}, volume 9207 of {\em Lecture
  Notes in Computer Science}, pages 163--179. Springer, 2015.

\bibitem{isola}
J.~Bessai, A.~Dudenhefner, B.~D{\"{u}}dder, M.~Martens, and J.~Rehof.
\newblock Combinatory logic synthesizer.
\newblock In T.~Margaria and B.~Steffen, editors, {\em Leveraging Applications
  of Formal Methods, Verification and Validation. Technologies for Mastering
  Change - 6th International Symposium, ISoLA 2014, Imperial, Corfu, Greece,
  October 8-11, 2014, Proceedings, Part {I}}, volume 8802 of {\em Lecture Notes
  in Computer Science}, pages 26--40. Springer, 2014.

\bibitem{comon:hal-03367725}
H.~Comon, M.~Dauchet, R.~Gilleron, F.~Jacquemard, D.~Lugiez, C.~L{\"o}ding,
  S.~Tison, and M.~Tommasi.
\newblock {\em {Tree Automata Techniques and Applications}}.
\newblock 2008.

\bibitem{DBLP:journals/scp/ErnstPGMPTX07}
M.~D. Ernst, J.~H. Perkins, P.~J. Guo, S.~McCamant, C.~Pacheco, M.~S. Tschantz,
  and C.~Xiao.
\newblock The {Daikon} system for dynamic detection of likely invariants.
\newblock {\em Sci. Comput. Program.}, 69(1-3):35--45, 2007.

\bibitem{DBLP:conf/popl/FengM0DR17}
Y.~Feng, R.~Martins, Y.~Wang, I.~Dillig, and T.~W. Reps.
\newblock Component-based synthesis for complex {APIs}.
\newblock In {\em Proceedings of the 44th {ACM} {SIGPLAN} Symposium on
  Principles of Programming Languages, {POPL} 2017, Paris, France, January
  18-20, 2017}, pages 599--612, 2017.

\bibitem{DBLP:conf/fm/FlanaganL01}
C.~Flanagan and K.~R.~M. Leino.
\newblock Houdini, an annotation assistant for {ESC/Java}.
\newblock In J.~N. Oliveira and P.~Zave, editors, {\em {FME} 2001: Formal
  Methods for Increasing Software Productivity, International Symposium of
  Formal Methods Europe, Berlin, Germany, March 12-16, 2001, Proceedings},
  volume 2021 of {\em Lecture Notes in Computer Science}, pages 500--517.
  Springer, 2001.

\bibitem{CGPRDS:GTWW75}
J.~Goguen, J.~Thatcher, E.~Wagner, and J.~Wright.
\newblock Abstract data-types as initial algebras and correctness of data
  representations.
\newblock In {\em Proceedings Conference on Computer Graphics, Pattern
  Recognition and Data Structure}, May 1975.

\bibitem{guo22}
Z.~Guo, D.~Cao, D.~Tjong, J.~Yang, C.~Schlesinger, and N.~Polikarpova.
\newblock Type-directed program synthesis for restful apis.
\newblock In R.~Jhala and I.~Dillig, editors, {\em {PLDI} '22: 43rd {ACM}
  {SIGPLAN} International Conference on Programming Language Design and
  Implementation, San Diego, CA, USA, June 13 - 17, 2022}, pages 122--136.
  {ACM}, 2022.

\bibitem{Thesis:Guttag75}
J.~V. Guttag.
\newblock {\em The Specification and Application to Programming of Abstract
  Data Types}.
\newblock PhD thesis, Computer Systems Research Group, Univ.\ of Toronto,
  Toronto, Canada, Sept. 1975.

\bibitem{DBLP:journals/acta/GuttagH78}
J.~V. Guttag and J.~J. Horning.
\newblock The algebraic specification of abstract data types.
\newblock {\em Acta Informatica}, 10:27--52, 1978.

\bibitem{DBLP:journals/tse/HenkelRD07}
J.~Henkel, C.~Reichenbach, and A.~Diwan.
\newblock Discovering documentation for {Java} container classes.
\newblock {\em {IEEE} Trans. Software Eng.}, 33(8):526--543, 2007.

\bibitem{DBLP:journals/ipl/HoodM81}
R.~Hood and R.~Melville.
\newblock Real-time queue operation in pure {LISP}.
\newblock {\em Inf. Process. Lett.}, 13(2):50--54, 1981.

\bibitem{DBLP:journals/tse/LiskovZ75}
B.~H. Liskov and S.~N. Zilles.
\newblock Specification techniques for data abstractions.
\newblock {\em {IEEE} Trans. Software Eng.}, 1(1):7--19, 1975.

\bibitem{DBLP:journals/pacmpl/MarianoRXNQFS19}
B.~Mariano, J.~Reese, S.~Xu, T.~Nguyen, X.~Qiu, J.~S. Foster, and
  A.~Solar{-}Lezama.
\newblock Program synthesis with algebraic library specifications.
\newblock {\em Proc. {ACM} Program. Lang.}, 3({OOPSLA}):132:1--132:25, 2019.

\bibitem{https://doi.org/10.48550/arxiv.2301.11117}
K.~Park, L.~D'Antoni, and T.~Reps.
\newblock Synthesizing specifications.
\newblock {\em CoRR}, abs/2301.11117, 2023.

\bibitem{CACM:Parnas72}
D.~L. Parnas.
\newblock On the criteria to be used in decomposing systems into modules.
\newblock {\em Comm.\ ACM}, 15(12):1053--1058, 1972.

\bibitem{ijcai15gulwani}
M.~Raza, S.~Gulwani, and N.~Milic-Frayling.
\newblock Compositional program synthesis from natural language and examples.
\newblock In {\em Proceedings of the 24th International Conference on
  Artificial Intelligence}, IJCAI'15, page 792–800. AAAI Press, 2015.

\bibitem{DBLP:journals/pacmpl/ShiSL19}
K.~Shi, J.~Steinhardt, and P.~Liang.
\newblock {FrAngel}: {C}omponent-based synthesis with control structures.
\newblock {\em Proc. {ACM} Program. Lang.}, 3({POPL}):73:1--73:29, 2019.

\bibitem{Simon:CeilingFloor73}
P.~Simon.
\newblock One man's ceiling is another man's floor, May 1973.
\newblock T-700.050.850-1 BMI, ISWC, JASRAC.

\bibitem{singh14}
R.~Singh, R.~Singh, Z.~Xu, R.~Krosnick, and A.~Solar{-}Lezama.
\newblock Modular synthesis of sketches using models.
\newblock In K.~L. McMillan and X.~Rival, editors, {\em Verification, Model
  Checking, and Abstract Interpretation - 15th International Conference,
  {VMCAI} 2014, San Diego, CA, USA, January 19-21, 2014, Proceedings}, volume
  8318 of {\em Lecture Notes in Computer Science}, pages 395--414. Springer,
  2014.

\bibitem{DBLP:journals/corr/abs-1905-06847}
J.~L. Singleton, G.~T. Leavens, H.~Rajan, and D.~R. Cok.
\newblock Inferring concise specifications of {APIs}.
\newblock {\em CoRR}, abs/1905.06847, 2019.

\bibitem{DBLP:journals/sttt/Solar-Lezama13}
A.~Solar{-}Lezama.
\newblock Program sketching.
\newblock {\em Int. J. Softw. Tools Technol. Transf.}, 15(5-6):475--495, 2013.

\bibitem{DBLP:journals/acta/SpitzenW74}
J.~M. Spitzen and B.~Wegbreit.
\newblock The verification and synthesis of data structures.
\newblock {\em Acta Informatica}, 4:127--144, 1974.

\bibitem{DBLP:journals/pacmpl/WillseyNWFTP21}
M.~Willsey, C.~Nandi, Y.~R. Wang, O.~Flatt, Z.~Tatlock, and P.~Panchekha.
\newblock egg: {F}ast and extensible equality saturation.
\newblock {\em Proc. {ACM} Program. Lang.}, 5({POPL}):1--29, 2021.

\end{thebibliography}
